\documentclass[%
 reprint,
nofootinbib,
 amsmath,amssymb,
 aps,showkeys,
]{revtex4-1}

\usepackage{graphicx}
\usepackage{dcolumn}
\usepackage{bm}
\usepackage{textpos}
\usepackage{hyperref}
\usepackage{amsmath}
\usepackage{breqn}

\begin{document}

\widetext

\preprint{APS/123-QED}

\begin{textblock*}{100mm}(-0.35cm,-1.1cm)
Submitted: 7 Jun 2017
\end{textblock*}

\begin{textblock*}{100mm}(.37\textwidth,-1.1cm)
Accepted: 28 Aug 2017
\end{textblock*}

\begin{textblock*}{100mm}(.73\textwidth,-1.1cm)
Published Online: 2 Oct 2017
\end{textblock*}

\begin{textblock*}{100mm}(.22\textwidth,-0.7cm)
Proc Natl Acad Sci USA (2017) vol. 114, no. xx, pg. xx
\end{textblock*}


\makeatletter
\providecommand\add@text{}
\newcommand\tagaddtext[1]{%
  \gdef\add@text{#1\gdef\add@text{}}}%
\renewcommand\tagform@[1]{%
  \maketag@@@{\llap{\add@text\quad}(\ignorespaces#1\unskip\@@italiccorr)}%
}
\makeatother

\newcommand{\beginsupplement}{%
        \setcounter{table}{0}
        \renewcommand{\thetable}{S\arabic{table}}%
        \setcounter{figure}{0}
        \renewcommand{\thefigure}{S\arabic{figure}}%
        \setcounter{equation}{0}
        \renewcommand{\theequation}{S\arabic{equation}}%
     }
     
\renewcommand{\thetable}{\arabic{table}}

\title{Origin of the RNA World: The Fate of Nucleobases in Warm Little Ponds}

\author{Ben K. D. Pearce\textsuperscript{1,4}}
\author{Ralph E. Pudritz\textsuperscript{1,2,3}} 
\author{Dmitry A. Semenov\textsuperscript{2}}
\author{Thomas K. Henning\textsuperscript{2}}

\footnotetext[1]{Origins Institute and Department of Physics and Astronomy, McMaster University, ABB 241, 1280 Main St, Hamilton, ON, L8S 4M1, Canada}
\footnotetext[2]{Max Planck Institute for Astronomy, K{\"o}nigstuhl 17, 69117 Heidelberg, Germany}
\footnotetext[3]{Center for Astronomy Heidelberg, Institute for Theoretical Astrophysics, Albert-Ueberle-Str. 2, 69120 Heidelberg, Germany}
\footnotetext[4]{To whom correspondence should be addressed. E-mail: pearcbe@mcmaster.ca}

\begin{abstract}
Prior to the origin of simple cellular life, the building blocks of RNA (nucleotides) had to form and polymerize in favourable environments on the early Earth. At this time, meteorites and interplanetary dust particles delivered organics such as nucleobases (the characteristic molecules of nucleotides) to warm little ponds whose wet-dry cycles promoted rapid polymerization. We build a comprehensive numerical model for the evolution of nucleobases in warm little ponds leading to the emergence of the first nucleotides and RNA. We couple Earth's early evolution with complex prebiotic chemistry in these environments. We find that RNA polymers must have emerged very quickly after the deposition of meteorites ($<$ a few years). Their constituent nucleobases were primarily meteoritic in origin and not from interplanetary dust particles. Ponds appeared as continents rose out of the early global ocean but this increasing availability of ``targets'' for meteorites was offset by declining meteorite bombardment rates. Moreover, the rapid losses of nucleobases to pond seepage during wet periods, and to UV photodissociation during dry periods means that the synthesis of nucleotides and their polymerization into RNA occurred in just one to a few wet-dry cycles. Under these conditions, RNA polymers likely appeared prior to 4.17 billion years ago.

\vspace{2mm}
{\it Significance:} There are currently two competing hypotheses for the site at
which an RNA world emerged: hydrothermal vents in the deep ocean and warm little ponds. Because the former lacks wet and dry cycles, which are well known to promote polymerization (in this case, of nucleotides into RNA), we construct a comprehensive model for the origin of RNA in the latter sites. Our model advances the story and timeline of the RNA world by constraining the source of biomolecules, the environmental conditions, the timescales of reaction, and the emergence of first RNA polymers.
\end{abstract}


\maketitle

One of the most fundamental questions in science is how life first emerged on the Earth.  Given its ubiquity in living cells and its ability to both store genetic information and catalyze its own replication, RNA probably formed the basis of first life \cite{1986Natur.319..618G}. RNA molecules are made up of sequences of 4 different nucleotides, the latter of which can be formed through reaction of a nucleobase with a ribose and a reduced phosphorous (P) source \cite{1963Natur.199..222P,Reference64}. The evidence suggests that first life appeared earlier than 3.7 Gyr ago (Ga) \cite{Reference134,Reference74} and thus the RNA world would have developed on a violent early Earth undergoing meteoritic bombardment at a rate of $\sim$1--1000 $\times$10$^{12}$ kg/yr \cite{1990Natur.343..129C}, which is approximately 8--11 orders of magnitude greater than today \cite{1996MNRAS.283..551B}. At this time, the atmosphere was dominated by volcanic gases, and dry land was scarce as continents were rising out of the global ocean. What was the source of the building blocks of RNA? And what environments enabled nucleotides to polymerize and form the first functioning RNA molecules under such conditions? Although experiments have produced simple RNA strands in highly idealized laboratory conditions \cite{Reference83,Reference58}, the answer to these questions are largely unknown.

As to the sources of nucleobases, the early Earth's atmosphere was likely dominated by CO$_2$, N$_2$, SO$_2$, and H$_2$O \cite{Reference119}. In such a weakly reducing atmosphere, Miller-Urey type reactions are not very efficient at producing organics \cite{1992Natur.355..125C}. One solution is that the nucleobases were delivered by interplanetary dust particles (IDPs) and meteorites. During these early times, these bodies delivered $\sim$6 $\times$10$^7$ kg yr$^{-1}$ and $\sim$2 $\times$10$^{3}$ kg yr$^{-1}$ of intact carbon, respectively \cite{1992Natur.355..125C}. Although nucleobases have not been identified in IDPs yet, three of the five nucleobases (uracil, cytosine, and thymine) have been formed on the surfaces of icy IDP analogues in the lab through exposure to UV radiation \cite{2014ApJ...793..125N}. Nucleobases are found in meteorites (guanine, adenine, and uracil) with concentrations of 0.25--515 parts-per-billion (ppb) \cite{Reference46,2011PNAS..10813995C}. The ultimate source of nitrogen within them could be molecules such as ammonia and HCN that are observed in the disks of gas and dust around all young stars \cite{Reference106,2016A&A...591A.122S,2009ApJ...696..143P}. A second possible source of nucleobases is synthesis in hydrothermal vents that well up from spreading cracks on the young Earth's ocean floors \cite{2001AsBio...1..133S}. A potential problem here is the lack of concentrated and reactive nitrogen sources in these environments \cite{2001AsBio...1..133S}.  

\begin{figure*}[hbtp]
\centering
\includegraphics[width=17.8cm]{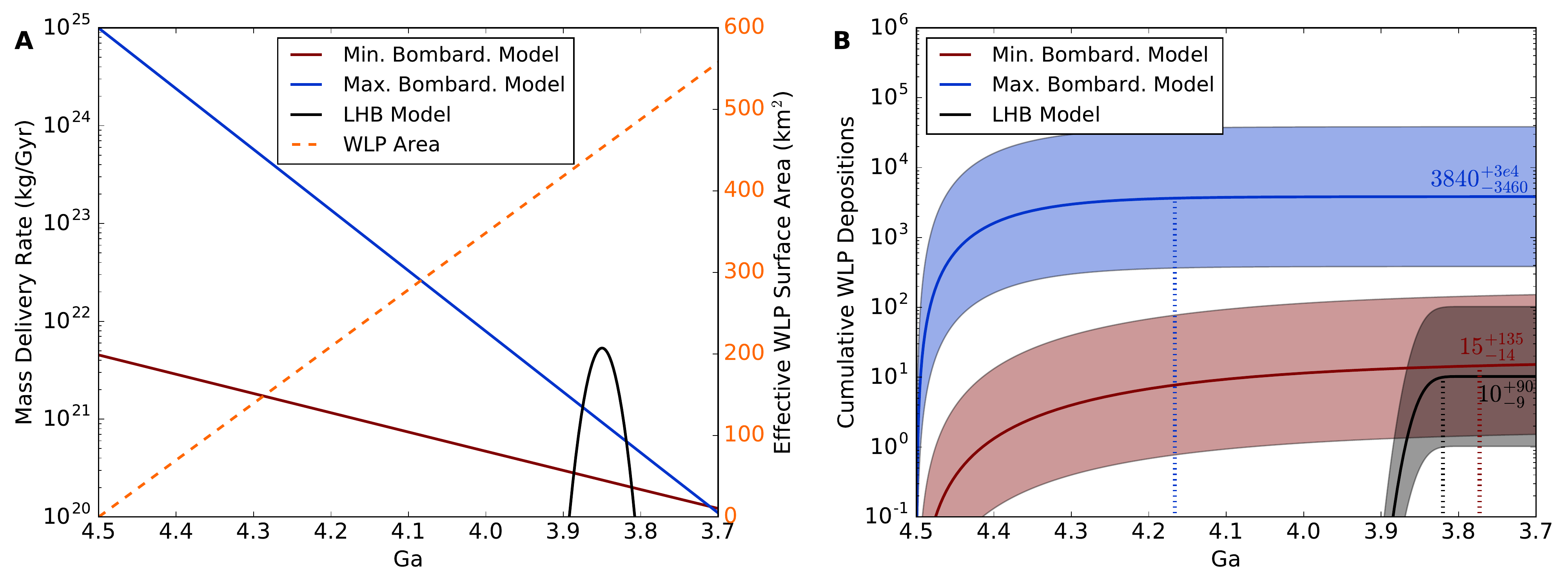}
\caption{{\bf History of carbonaceous meteorite deposition in warm little ponds: (A)} History of mass delivery rate to the early Earth and of effective warm little pond (WLP) surface area. Three models for mass delivered to the early Earth are compared: a Late Heavy Bombardment model, and a minimum and maximum bombardment model. All mass delivery models are based on analyses of the lunar cratering record \cite{2009Natur.459..419A,1990Natur.343..129C}. The effective WLP surface area during the Hadean eon is based on a continental crust growth model \cite{Reference136}, the number of lakes and ponds per unit crustal area \cite{Reference98}, and the lake and pond size distribution \cite{Reference98}. {\bf (B)} Cumulative WLP depositions for the small fragments of carbonaceous meteoroids of diameter 40--80 m landing in any 1--10 m-radius WLP on the early Earth. A deposition is characterized as a meteoroid debris field that overlaps with a WLP. We assume the ponds per area during the Hadean eon is the same as today, and that all continental crust remains above sea-level, however we vary the ponds per area by $\pm$1 order of magnitude to obtain error bars. 95$\%$ of WLP depositions in each model occur before the corresponding dotted vertical line. 40--80 m is the optimal range of carbonaceous meteoroid diameters to reach terminal velocity, while still landing a substantial fraction of mass within a WLP area. 1--10 m is the optimal range of WLP radii to avoid complete evaporation within a few months, while also allowing non-negligible nucleobase concentrations to be reached upon meteorite deposition. The numbers of carbonaceous meteoroid impactors in the 20--40 m-radius range are based on the mass delivery rate, the main-belt size-frequency distribution, and the fraction of impactors that are of type CM, CR or CI \cite{1990Natur.343..129C,2009Natur.459..419A,2002aste.book..653B} (see SI Text).}
\label{ProbResults}
\end{figure*}

A central question concerning the emergence of the RNA world is how the polymerization of nucleotides occurred. Warm little ponds (WLPs) are excellent candidate environments for this process because their wet and dry cycles have been shown to promote the polymerization of nucleotides into chains possibly greater than 300 links \cite{Reference83}. Furthermore, clay minerals in the walls and bases of WLPs promote the linking of chains up to 55 nucleotides long \cite{1996Natur.381...59F}. Conversely, experiments simulating the conditions of hydrothermal vents have only succeeded in producing RNA chains a few monomers long \cite{2015AsBio..15..509B}. A critical problem for polymerizing long RNA chains near hydrothermal vents is the absence of wet-dry cycles.

\subsection*{Model: Fates of Nucleobases in Evolving WLPs}

We compute a well posed model for the evolution of WLPs, the fates of nucleobases delivered to them, and the emergence of RNA polymers under early Earth conditions. The sources of nucleobases in our model are carbonaceous meteorites and IDPs whose delivery rates are estimated using the lunar cratering record \cite{1990Natur.343..129C,1992Natur.355..125C,2009Natur.459..419A}, the distribution of asteroid masses \cite{2009Natur.459..419A}, and the fraction of meteorites reaching terminal velocity that are known to be nucleobase carriers \cite{2002aste.book..653B,Reference46}. The WLPs are ``targets'' in which molecular evolution of nucleobases into nucleotides and subsequent polymerization into RNA occurs. The evolution of ponds due to precipitation, evaporation, and seepage constitute the immediate environments in which the delivered nucleobases must survive and polymerize\footnote{We don't emphasize groundwater-fed ponds (hot springs) because their small number on Earth today ($\sim$thousands) compared to regular lakes/ponds ($\sim$304 million \cite{Reference98}) suggests they didn't contribute greatly to the combined WLP target area for meteorite deposition.}. The data for these sources and sinks are gathered from historical precipitation records \cite{Reference113}, or are measured experimentally \cite{Reference99,Reference116}. Sinks for these molecules include destruction by unattenuated UV rays (as the early Earth had no ozone), hydrolysis in the pond water, as well as seepage from the bases of ponds. The data for these sinks are measured experimentally \cite{2015AsBio..15..221P,Reference45,Reference116}. The nucleobases that survive go on to form nucleotides that ultimately polymerize.

To calculate the number of carbonaceous meteorite source depositions in target WLPs on the early Earth, we combine a continental crustal growth model \cite{Reference136}, the lake and pond size distribution and ponds per unit crustal area estimated for ponds on Earth today \cite{Reference98}, the asteroid belt mass distribution \cite{2009Natur.459..419A} and three possible mass delivery models based on the lunar cratering record \cite{1990Natur.343..129C,2009Natur.459..419A}. This results in a first order linear differential equation (Equation~\ref{totalprob}), which we solve analytically. The details are in the SI Text, and the main results are discussed in the following section.

Nucleobase abundances in WLPs from IDPs and meteorites are described in our model by first order linear differential equations (Equations~\ref{IDPAccumEq} and \ref{MetAccumEq}, respectively). The equations are solved numerically (see SI Text for details).

Our fiducial model WLPs are cylindrical (because sedimentation flattens their initial bases) and have a 1 m radius and depth. We find these ponds to be optimal because they are large enough to not dry up too quickly but small enough that high nucleobase concentrations can be achieved (see SI Text).

\begin{figure*}[hbtp]
\centering
\includegraphics[width=11.4cm]{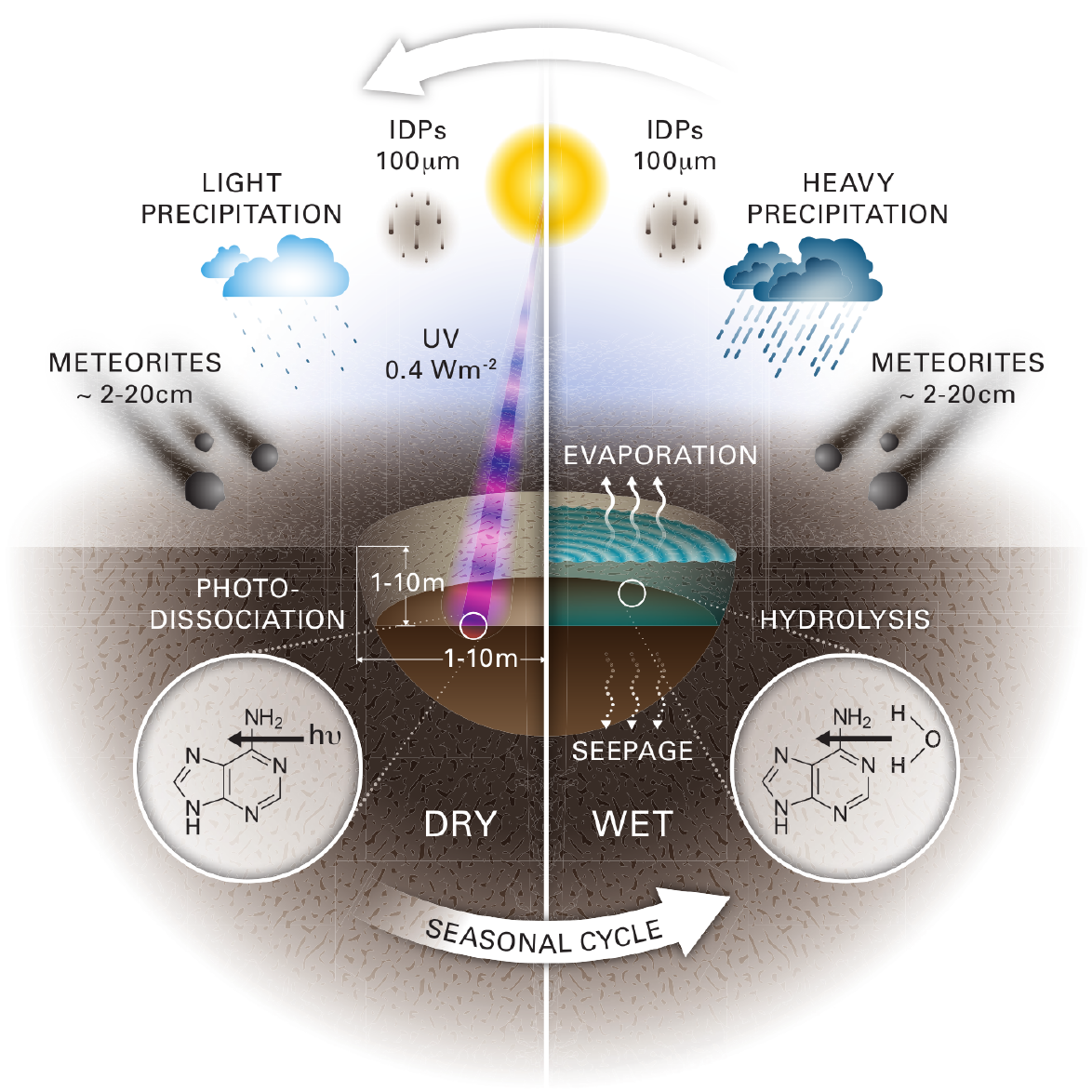}
\caption{An illustration of the sources and sinks of pond water and nucleobases in our model of isolated warm little ponds on the early Earth. The only water source is precipitation. Water sinks include evaporation and seepage. Nucleobase sources include carbonaceous IDPs and meteorites, which carry up to $\sim$1 picogram and $\sim$3 mg of each nucleobase, respectively. Nucleobase sinks include hydrolysis, UV photodissociation, and seepage. Nucleobase hydrolysis and seepage are only activated when the pond is wet, and UV photodissociation is only activated when the pond is dry.}
\label{WLPMainFig}
\end{figure*}

\subsection*{Meteorite Sources and Targets}

In Figure~\ref{ProbResults}A, we show the history of meteoritic mass delivery rates on the early Earth, and of WLP targets. From total meteoritic mass, we extract just the nucleobase sources, i.e. only carbonaceous meteoroids whose fragments slow to terminal velocity in the early atmosphere. In Figure~\ref{ProbResults}B we show the history of depositions of such nucleobase sources into 1--10 m radius WLP targets from 4.5--3.7 Ga. 

The lunar cratering record shows either a continuously decreasing rate of impacts from 4.5 Ga forwards, or a brief outburst beginning at $\sim$3.9 Ga known as the Late Heavy Bombardment (LHB). Due to the lack of lunar impact melts with ages prior to $\sim$3.92 Ga, the rate of mass delivered to the early Earth preceding this date is uncertain \cite{2007SSRv..129...35Z}. Therefore we compare three models for mass delivery to the early Earth based on analytical fits to the lunar cratering record \cite{2009Natur.459..419A,1990Natur.343..129C}: a LHB, a minimum and a maximum bombardment model. The latter two models represent lower and upper limit fits for a sustained declining bombardment, and are subject to considerable uncertainty \cite{1991Icar...92..217C}. The total target area of WLPs increases as the rising continents provide greater surface area for pond sites. An ocean world, in this view, is an implausible environment for the emergence of an RNA world. Iron and rocky bodies impact the solid surface at speeds high enough to form small craters \cite{1993AJ....105.1114H}; contributing to the WLP population.

Our calculations show that 10--3840 terminal velocity carbon-rich meteoroids would have deposited their fragments into WLPs on the Earth during the Hadean eon. Given the large uncertainty of the ponds per unit area and the growth rate of continental crust, we vary the WLP growth function by $\pm$ 1 order of magnitude. From this we get a minimum of 1--384 depositions and a maximum of 100--38,400 WLP depositions from 4.5--3.7 Ga. The optimal model for WLP depositions---the maximum bombardment model---suggests the majority of depositions occurred prior to 4.17 Ga. The less optimal models for WLP depositions, the LHB and minimum bombardment models, suggest the majority of depositions occurred before 3.82 Ga and 3.77 Ga, respectively.

\subsection*{Life Cycles of WLPs} 

Figure~\ref{WLPMainFig} illustrates the variation of physical conditions for WLPs during the Hadean eon, approximately 4.5--3.7 Ga. Annual rainfall varies sinusoidally \cite{Reference113}, creating seasonal wet and dry environments. The increased heat flow from greater abundances of radiogenic sources at this time \cite{Reference89} causes temperatures of around 50--80 $^{\circ}$C \cite{Reference80}. The various factors that control the water level and thus the wet-dry cycles of WLPs are precipitation, evaporation, and seepage (through pores in the ground). 

In Figure~\ref{IDPWater}A, we present the results of these calculations. We select two different temperatures (65$^{\circ}$C, and 20$^{\circ}$C) as analogues for hot and warm early Earths. For each of these, we examine three different environments: dry, intermediate, and wet (see Table~\ref{precipModels} for details).
The water levels in the wet environment WLPs range from approximately 60--100$\%$ full. WLPs experiencing dry states of approximately half (intermediate environment) and three-quarters of a year (dry environment) only fill up to 20$\%$ and 10$\%$, respectively. These results clearly establish the existence of seasonal wet-dry cycles.

\subsection*{Nucleobase Evolution in WLPs}

As shown in Figure~\ref{WLPMainFig}, the build up of nucleobases in WLPs is offset by losses due to hydrolysis \cite{Reference45}, seepage \cite{Reference116}, and dissociation by UV radiation that was incident on the early Earth in the absence of ozone \cite{2015AsBio..15..221P,2015ApJ...806..137R}. Some protection would be afforded during WLP wet phases, as a 1 m column of pond water can absorb UV radiation up to $\sim$95$\%$ \cite{Reference114}. It is of particular interest that sediment, which collects at the base of WLPs, also attenuates UV radiation. Studies show it only takes a $\sim$0.6 mm layer of basaltic sediment to attenuate UV radiation by $>$99.99$\%$ \cite{2017LPI....48.2678C}.  


\begin{figure*}[hbtp]
\centering
\includegraphics[width=17.8cm]{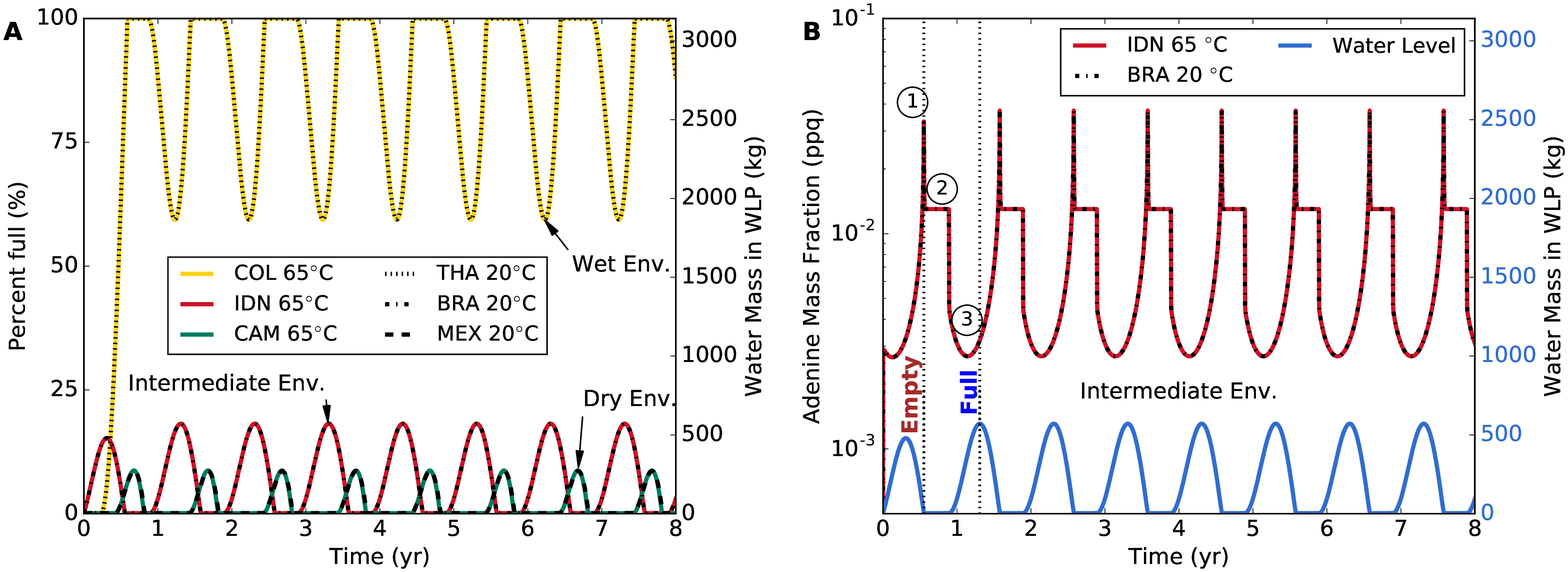}
\caption{ {\bf Histories of pond water and adenine concentration from interplanetary dust particles: (A)} The change in water level over time in our fiducial dry, intermediate, and wet environment WLPs due to evaporation, seepage, and precipitation. Precipitation rates from a variety of locations on Earth today are used in the models, and represent 2 classes of matching early Earth analogues: hot (Columbia, Indonesia, Cameroon), and warm (Thailand, Brazil, and Mexico) (for details see Table~\ref{precipModels}). All models begin with an empty pond, and stabilize within 2 years. {\bf (B)} The red and black-dotted curves represent the adenine concentrations over time from carbonaceous interplanetary dust particles (IDPs) in our fiducial WLPs. The degenerate intermediate WLP environment used in this calculation is for a hot early Earth at 65 $^{\circ}$C and a warm early Earth at 20 $^{\circ}$C. The blue curve represents the corresponding water level in the WLP, with initial empty and full states labeled vertically. Three features are present: 1) The maximum adenine concentration at the onset of the dry phase. 2) A flat-top equilibrium between incoming adenine from IDPs and adenine destruction by UV irradiation. 3) The minimum adenine concentration just before the pond water reaches its highest level.}
\label{IDPWater}
\end{figure*}

Nucleotide formation and stability are sensitive to temperature. Phosphorylation of nucleosides in the lab is slower at low temperatures, taking a few weeks at 65$^{\circ}$C compared to a couple hours at 100$^{\circ}$C \cite{Reference300}. The stability of nucleotides on the other hand, is favoured in warm conditions over high temperatures \cite{Reference125}. If a WLP is too hot ($>$80$^{\circ}$C), any newly formed nucleotides within it will hydrolyze in several days to a few years \cite{Reference125}. At temperatures of 5--35 $^{\circ}$C that either characterize more temperate latitudes or a post-snowball Earth, nucleotides can survive for thousand-to-million-year timescales. However at such temperatures, nucleotide formation would be very slow. Considering this temperature sensitivity, we model the evolution of nucleobases in WLPs in matching warm (5--35 $^{\circ}$C) and hot (50--80 $^{\circ}$C) early Earth environments. Hotter ponds evaporate quicker, therefore we choose rainier analogue sites on a hot early Earth to maintain identical pond environments to our warm early Earth sites.

\begin{table*}[!ht]
\centering
\caption{Precipitation models matching dry, intermediate, and wet environments on a warm (5--35$^{\circ}$C) and hot (50--80$^{\circ}$C) early Earth. Precipitation data from a variety of locations on Earth today \cite{Reference113,Reference117} represent 2 classes of matching early Earth analogues: warm (Thailand, Brazil, and Mexico) and hot (Columbia, Indonesia, Cameroon). For example, the conditions in Mexico on a warm early Earth match the conditions in Cameroon on a hot early Earth. $\overline{P}$ is the mean precipitation rate, $\delta_p$ is the seasonal precipitation amplitude, and $s_p$ is the phase shift. \label{precipModels}}

\begin{tabular}{cccccc}
\\
\multicolumn{1}{c}{Model} &
\multicolumn{1}{c}{Environment} &
\multicolumn{1}{c}{Analogue Site} &
\multicolumn{1}{c}{$\overline{P}$ (m yr$^{-1}$)} &
\multicolumn{1}{c}{$\delta_p$} & 
\multicolumn{1}{c}{$s_p$ (yr)}\\ \hline \\[-2.5mm]
Warm early Earth (5--35$^{\circ}$C) & Dry & Mexico (MEX) & 0.94 & 1.69 & 0.3\\
& Intermediate & Brazil (BRA) & 1.8 & 0.50 & 0.85\\
& Wet & Thailand (THA) & 3.32 & 0.91 & 0.3\\ \hline \\[-2.5mm]
Hot early Earth (50--80$^{\circ}$C) & Dry & Cameroon (CAM) & 3.5 & 0.5 & 0.3\\
& Intermediate & Indonesia (IDN) & 4.5 & 0.2 & 0.85\\
& Wet & Columbia (COL) & 6.0 & 0.5 & 0.3\\ 
\hline
\multicolumn{6}{l}{\footnotesize To obtain the rate of the decrease in pond water for a given analogue site, table values are input into this equation: }\\
\multicolumn{6}{l}{\footnotesize $\frac{dL}{dt} = 0.83 + 0.06T - \overline{P} \left[ 1 + \delta_p sin \left( \frac{2 \pi (t - s_p )}{\tau_s} \right) \right]$ (see SI Text).}
\end{tabular}
\end{table*}

In Figure~\ref{IDPWater}B, we focus on the adenine concentrations in WLPs from only IDP sources. The combination of spikes, flat tops, and troughs in Figure~\ref{IDPWater}B reflects the variations of adenine concentration in response to drying, balance of input and destruction rates, and precipitation during wet periods. In any environment and at any modeled temperature, the maximum adenine concentration from only IDP sources remains below 2 $\times$ 10$^{-7}$ppb (see Figure~\ref{IDPAccum}). This is two orders of magnitude below current detection limits \cite{2011PNAS..10813995C}, making subsequent reactions negligible. The nucleobase mass fraction curves are practically independent of pond size (1m $<$ r$_p$ $\leq$ 10 m) once a stable, seasonal pattern is reached ($<$ 35 years). This is because although larger ponds have a larger collecting area for IDPs, they have an equivalent larger area for collecting precipitation and seeping nucleobases.

\subsection*{Dominant Source of Surviving Nucleobases} 

\begin{figure*}[hbtp]
\centering
\includegraphics[width=17.8cm]{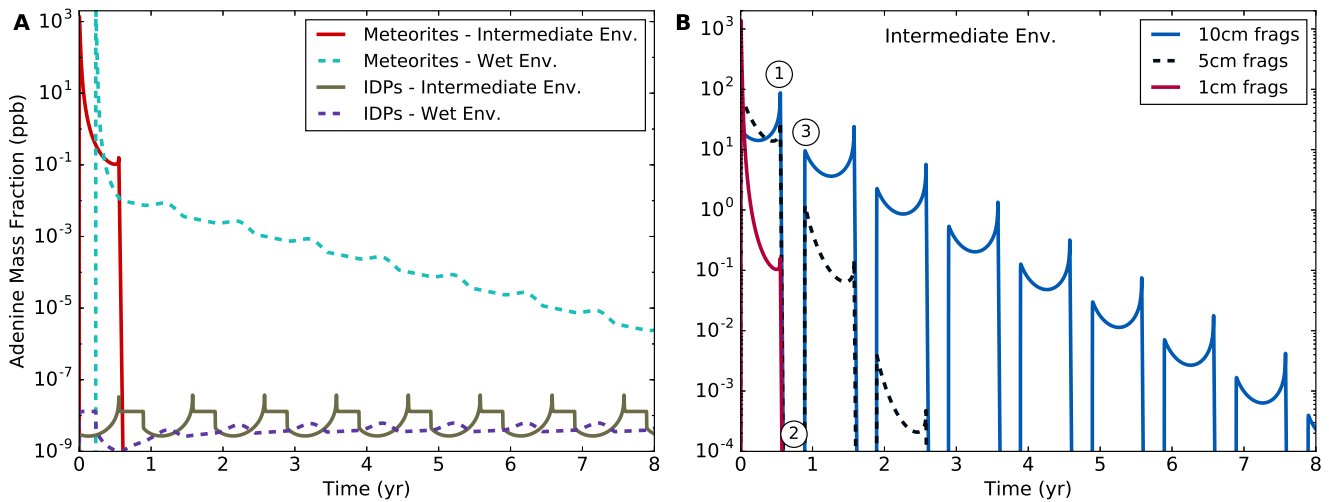}
\caption{{\bf Comparative histories of adenine concentrations from interplanetary dust particles and meteorites: (A)} A comparison of the accumulation of adenine from carbonaceous IDPs and meteorites in our fiducial WLPs. The meteorite fragments are small (1 cm), and originate from a 40 m-radius carbonaceous meteoroid. Adenine concentrations for intermediate (wet-dry cycle) and wet environments (never dry) are compared and correspond to both a hot early Earth at 65 $^{\circ}$C and a warm early Earth at 20 $^{\circ}$C (for details see Table~\ref{precipModels}). {\bf (B)} The effect of meteorite fragment sizes on adenine concentration. The degenerate intermediate WLP environment used in these calculations is for a hot early Earth at 65 $^{\circ}$C and a warm early Earth at 20 $^{\circ}$C. The fragments are either only small in size (1 cm in radius), only medium in size (5 cm in radius), or only large in size (10 cm in radius). Three features are present: 1) Adenine is at its highest concentration at the onset of the pond's dry phase. 2) Upon drying, adenine ceases to outflow from large fragments, and UV radiation rapidly destroys all previously released adenine. 3) Re-wetting allows the remaining adenine within large fragment pores to continue to outflow. The U-shape is due to the increase and decrease in water level.}
\label{FinalFigure}
\end{figure*}

In Figure~\ref{FinalFigure} we assemble all of these results and compare carbonaceous IDPs to meteorites as sources of adenine to our fiducial WLPs. Small meteorite fragments (1 cm in radius) are compared with IDPs in panel A. The effects of larger meteorite fragments (5 cm and 10 cm) on adenine concentration are displayed in panel B. 

The maximum adenine concentration in our model WLPs from carbonaceous meteorites is 10 orders of magnitude higher than the maximum adenine concentration from carbonaceous IDPs. The reason for this huge disparity is simply that carbonaceous meteoroid fragments---each carrying up to a few mg of adenine---are deposited into a WLP in a \emph{single event}. This allows adenine to reach ppb--ppm-level concentrations before seepage and UV photodissociation efficiently remove it from the WLP in one to a few wet-dry cycles. The maximum guanine and uracil accumulated in our model WLPs from meteorites are also more than 10 orders of magnitude higher than those accumulated from IDPs (see Figures~\ref{IDPAccum65All} and \ref{MetAccum65All}). A maximum adenine concentration of 2 ppm is still $\sim$1--2 orders of magnitude lower than the initial adenine concentrations in aqueous experiments forming adenosine and AMP \cite{1963Natur.199..222P}, however these experiments only ran for an hour. 

Adenine within larger fragments diffuses over several wet-dry cycles, and during the dry phase no outflow occurs. For fragments 1, 5, and 10 cm in radius, 99$\%$ of the adenine is released into the pond water within $\sim$10 days, 8 months, and 32 months, respectively (see Figure~\ref{IDPMetDiff}). Only the adenine that has already flowed out of the fragments gets rapidly photodestroyed, therefore, adenine within larger fragments can survive up to $\sim$7 years.

Thus, even though the carbon delivery rates from IDPs to the early Earth vastly exceed those from meteorites, it is the meteoritic material that is the dominant source of nucleobases for RNA synthesis. 

\subsection*{Nucleotide and RNA Synthesis}

To form nucleotides in WLPs, ribose and a reduced P source must be available. Ribose may have formed and stabilized in borate-rich WLPs via the formose reaction (polymerization of H$_2$CO) \cite{Reference301}. Additionally, phosphite has been detected in a pristine geothermal pool representative of early earth, suggesting the potential availability of reduced P to WLP environments \cite{Reference302}. Only the AMP nucleotide has been experimentally synthesized in a single step involving ribose, reduced P and UV radiation \cite{1963Natur.199..222P}.

Because of the rapid rate of seepage ($\sim$1.0--5.1 mm day$^{-1}$ \cite{Reference116}), nucleotide synthesis would need to be fast, occurring within a half-year to a few years after nucleobase deposition, depending on meteoroid fragment sizes. This is ample time given that lab experiments show hour-to-week-long timescales are sufficient to form adenosine and AMP \cite{1963Natur.199..222P,Reference64}. 

Nucleotides, once synthesized using meteorite-delivered nucleobases, are still subject to seepage, regardless of the temperature. Therefore nucleotide polymerization into RNA would also need to be fast, occurring within one to a few wet-dry cycles, in order to reduce their likelihood of seeping through the estimated 0.001--400 $\mu$m-sized pores at the WLP base \cite{Reference133}. Experiments show that nucleotides can polymerize into RNA chains possibly greater than 300 monomers by subjecting them to just 1--16 wet-dry cycles \cite{Reference83}.

\subsection*{Discussion and Conclusions}

Seepage is one of the dominant nucleobase sinks in WLPs. It will be drastically reduced if nucleobases are encapsulated by vesicles (spheres of size 0.5-5 $\mu$m that form spontaneously upon hydration, whose walls consist of lipid bilayers derived from fatty acids within meteorites) \cite{Reference87}. However, even if seepage is turned off, maximum adenine concentrations from IDPs are still negligible ($\lesssim$150 parts per quadrillion) (see Figure~\ref{IDPAccumNoSeepage}).

Also, we note that a cold early Earth (if it occurred \cite{2007SSRv..129...35Z}) with seasonal or impact-induced freeze-thaw cycles could also be suitable for RNA polymerization and evolution for reasons similar to those we have analyzed for WLPs. The cyclic thawing and freezing resembles wet-dry cycles \cite{Reference83}.

We conclude that the physical and chemical conditions of WLPs place strong constraints on the emergence of an RNA world. A hot early Earth (50--80$^{\circ}$C) favours rapid nucleotide synthesis in WLPs \cite{Reference300}. Meteorite-delivered nucleobases could react with ribose and a reduced P source to quickly ($<$ a few years) create nucleotides for polymerization. Polymerization then occurs in one to a few wet-dry cycles to reduce the likelihood that these molecules are lost to seepage. This rapid process also reduces the likelihood of setbacks for the emergence of the RNA world due to frequent large impacts, a.k.a. impact frustration \cite{2006coel.conf..207Z}. Sedimentation would be of critical importance as UV protection for nucleobases, nucleotides, and RNA \cite{2015AsBio..15..221P,Reference130,2012NatSR...2E.517K}.

The mass-delivery model providing the most WLP depositions indicates that the majority of meteorite depositions occurred prior to 4.17 Ga. The first RNA polymers would have formed in WLPs around that time, prefiguring the emergence of the RNA world. This implies that the RNA world could have appeared within $\sim$200--300 million years after the Earth became habitable.

\subsection*{Acknowledgements}

We thank the referees for their thoughtful insights. We thank M. Rheinst{\"a}dter for providing valuable comments on the manuscript. B.K.D.P. thanks the Max Planck Institute for Astronomy for their hospitality during his summer abroad, supported by a Natural Sciences and Engineering Research Council of Canada (NSERC) Michael Smith Foreign Study Supplement. R.E.P. thanks the Max Planck Institute for Astronomy and the Institute of Theoretical Astrophysics for their support during his sabbatical leave. The research of B.K.D.P. was supported by an NSERC Canada Graduate Scholarship and Ontario Graduate Scholarship. R.E.P. is supported by an NSERC Discovery Grant. This research is part of a collaboration between the Origins Institute and the Heidelberg Initiative for the Origins of Life.

\subsection*{Author Disclosure Statement}

The authors declare no conflict of interest.


\bibliography{Bibliography_MetIDPForPNAS}

\section*{Supporting Information (SI)}

\beginsupplement

\subsection*{Calculating WLP Surface Area on the Early Earth}

The earliest fairly conclusive evidences of life are in the form of light carbon signatures in graphite globules formed from marine sediments, and stromatolite fossils, both which are dated to be 3.7 Gyr old \cite{Reference74,Reference134}. Therefore nucleobase deposition in WLPs and subsequent reactions to form nucleotides and RNA, would have occurred sometime between $\sim$4.5--3.7 Ga. 

Once Theia impacted the proto-Earth $\sim$50 Myr after Solar System formation, it may have only taken 10--100 Myr for the magma mantle to cool \cite{2007SSRv..129...35Z}. Furthermore, basic evidence (from the zircon oxygen isotope record) exists of crustal material interacting with liquid water at or near the Earth's surface since 4.3 Ga \cite{Reference90}. Such evidence makes it clear that the Earth had a hydrosphere at that point, and therefore could have formed WLPs on any rising continental crust. 

An estimate of the surface area of WLPs on the early Earth has not been previously attempted, however it is often suggested that they were typical on early Earth continents \cite{Reference87,Reference89,Reference90}. Water-deposited sediments dated at 3.8 Ga indicate early erosion and transport of sediment, therefore at least some continental mass must have been exposed above sea level at that time, on which WLPs could have formed \cite{Reference89}. 

\subsubsection*{Continental Crust Growth Model}

The number of WLPs present at any given time depends on the fraction of continental crust above sea level. Calculations from a continental crust growth model shows a linear formation of early crust, increasing from 0 to 12.8$\%$ of today's crustal surface area from 4.5--3.7 Ga \cite{Reference136}. This is expressed in the equation below.

\begin{equation}\label{fracCrust}
f_{cr} = 0.16t
\end{equation}
where $f_{cr}$ is the fraction of today's crustal surface area, and t is the time in Gyr ($t$ = 0 starts at 4.5 Ga).

We assume the number of bodies of water per unit area of continental crust is constant over time, thus by multiplying Equation \ref{fracCrust} by the number of lakes and ponds on Earth today, we get the number of bodies of water on Earth at any date from 4.5--3.7 Ga. The number of lakes and ponds on Earth today (down to 0.001 km$^2$) is estimated to be 304 million \cite{Reference98}, therefore the equation becomes:

\begin{equation}\label{NumPonds}
N_t = 4.864 \times 10^{7} t
\end{equation}
where $N_t$ is the total number of bodies of water on Earth for times 0 Gyr $\leq t \leq$ 0.8 Gyr ($t$ = 0 starts at 4.5 Ga).

\subsubsection*{Lake and Pond Size Distribution}

The size distribution of lakes and ponds on Earth today follows a Pareto distribution function \cite{Reference98}.

\begin{equation}\label{pareto}
dN_A = N_t c k^c A^{-(c + 1)} dA,
\end{equation}
where $dN_A$ is the number of bodies of water in the km$^2$ area range $A$ to $A + dA$, $N_t$ is the total number of bodies of water on Earth, $c$ is the dimensionless shape parameter, and k is the smallest lake/pond area in the distribution. The shape parameter was calculated for lakes and ponds down to 0.001 km$^{2}$ to be $c$ = 1.06 \cite{Reference98}.

The total area of ponds on Earth in a given size range can then be calculated by multiplying Equation~\ref{pareto} by $A$ and integrating from $A_{min}$ to $A_{max}$, which gives

\begin{equation}\label{pareto2}
\hspace{9mm}
A_{tot} = \frac{c}{-c+1} N_t k^c \left[A_{max}^{(-c + 1)} - A_{min}^{(-c + 1)}\right]
\hspace{3mm}
\textrm{[km$^2$].}
\end{equation}

There is an upper size limit on WLPs in which substantial concentrations of nucleobases from meteorites can be deposited. If the surface area of a WLP is comparable to the area of a meteoroid's strewnfield, then partial overlapping of the strewnfield and WLP is probable. This would lead to fewer fragment depositions per unit pond area, and lower nucleobase concentrations. For this reason, we choose the upper limit on WLP radii to be 10 m. This equates to pond areas that are $<$0.04 $\%$ of the strewnfield area of typical carbonaceous meteoroids (the latter of which are $\sim$0.785 km$^2$).

There is also a lower size limit as ponds that evaporate too quickly spend the majority of their time in the dry state. This would prevent nucleobase outflow from the pores of meteorites. Moreover, the probability of meteorite deposition in WLPs decreases for decreasing pond radii. A cylindrical WLP at 65 $^{\circ}$C with a radius and depth of $<$ 1 metre will likely evaporate in less than 3 months without replenishment from rain or geysers \cite{Reference99}. Therefore we set the lower WLP radius and depth limit for our interests at 1 m. 

\subsubsection*{Total WLP Surface Area}

From Equations~\ref{NumPonds} and \ref{pareto2}, the total surface area of cylindrical, 1--10 m-radii WLPs on the early Earth at time 0 Gyr $\leq t \leq$ 0.8 Gyr is

\begin{equation}\label{TotArea}
A_{tot}(t) = 651 t
\tagaddtext{[km$^2$].}
\end{equation}
For this calculation, we choose $c$ = 1.06 and $k$ = 3.14$\times$10$^{-6}$ km$^2$. Note $t$ = 0 starts at 4.5 Ga.


\subsection*{Calculating Carbonaceous Meteorite Depositions in WLPs}\label{AppendixA}

Aerodynamic forces fragment meteoroids that enter the atmosphere, which increases their total meteoroid cross-section, and thus their aerodynamic braking \cite{1993AJ....105.1114H}. Numerical simulations show that the fragments of carbonaceous meteoroids with initial diameters up to 80 metres and atmospheric entry velocities near the median value (15 km/s \cite{1991Icar...92..217C}) reach terminal velocity \cite{1993AJ....105.1114H}. These fragments do not produce craters upon impact \cite{2009M&PS...44..985K} and can be intactly deposited into WLPs. Larger meteoroids produce fragments with impact speeds too fast to avoid cratering and partial or complete melting or vapourization of the impactor. The same numerical simulations calculate the largest fragments from a 80-metre-diameter carbonaceous meteoroid of initial velocity 15 km/s to be $\sim$20 cm in diameter \cite{1993AJ....105.1114H}. (This roughly makes sense as the biggest carbonaceous meteorite recovered, the Allende, is only 110 kg---corresponding to a $\sim$ 40 cm-diameter sphere \cite{1971SmCES...5....1C}). 

The optimal diameter range for carbonaceous meteoroids to deposit a substantial fraction of mass into WLPs at terminal velocity is therefore 40--80 m. We base our calculation of carbonaceous meteoroid fragment depositions on this range.

\subsubsection*{Mass Delivery Models}

The lunar cratering record analyzed by the Apollo program has revealed a period of intense lunar bombardment from $\sim$3.9--3.8 Ga \cite{2005Sci...309.1847S}. Whether a single lunar cataclysm (lasting $\sim$10--150 Myr) or a sustained declining bombardment preceded 3.9 Ga is still debated \cite{2007SSRv..129...35Z}. We choose three models for the rate of mass delivered to the early Earth: a LHB model, and a minimum and maximum bombardment model. All models are based on analyses of the lunar cratering record \cite{2009Natur.459..419A,1990Natur.343..129C}.

Analyses from both dynamic modeling and the lunar cratering record estimate the total mass delivered to the early Earth during the LHB to be $\sim 2 \times 10^{20}$ kg \cite{2009Natur.459..419A}. We assume a Gaussian curve for the rate of impacts during the LHB \cite{2007SSRv..129...35Z}, which centers on 3.85 Ga, integrates to $2 \times 10^{20}$ kg, and drops to nearly zero at 3.9 and 3.8 Ga. Thus,

\begin{dmath}\label{LHBmass}
dm_{LHB}(t) = 5.33 \times 10^{21} e^{\frac{-(t-0.65)^2}{2(0.015)^2}} dt
\hspace{3mm}
\textrm{[kg],}
\end{dmath}
where $dm_{LHB}$ is the total mass from $t$ to $t + dt$ (Gyr) ($t$ = 0 starts at 4.5 Ga).

Equations for the minimum and maximum mass delivered to the early Earth, given a sustained declining bombardment preceded 3.9 Ga, are displayed below \cite{1990Natur.343..129C}.

\begin{dmath}\label{minmass}
m_{minB}(t) = 1 \times 10^{21} - 0.76\left[4.5 - t + 4.57 \times 10^{-7} \left( e^{(4.5 - t)/\tau_a} - 1 \right) \right] m_2^{1 - b} 4 \pi R_{moon}^2 \Sigma
\hspace{3mm}
\textrm{[kg].}
\end{dmath}

\begin{dmath}\label{maxmass}
m_{maxB}(t) = 7 \times 10^{23} - 0.4\left[4.5 - t + 5.6 \times 10^{-23} \left( e^{(4.5 - t)/\tau_c} - 1 \right) \right] m_2^{1 - b} 4 \pi R_{moon}^2 \Sigma
\hspace{3mm}
\textrm{[kg].}
\end{dmath}
where $m_{minB}$ and $m_{maxB}$ are the total mass delivered from $t$ = 0 to $t = t$ ($t$ = 0 starts at 4.5 Ga), $\tau_a$ and $\tau_b$ are decay constants ($\tau_a$ = 0.22 Gyr, $\tau_b$ = 0.07 Gyr), $m_2$ is the maximum impactor mass ($m_2$ = 1.5 $\times$ 10$^{18}$ kg), b = 0.47, $R_{moon}$ is the mean radius of the moon ($R_{moon}$ = 1737.1 km), and $\Sigma$ is the ratio of the gravitational cross sections of the Earth and Moon ($\Sigma$ $\sim$23).

Taking the derivatives of Equations~\ref{minmass} and \ref{maxmass} gives us the corresponding rates of mass delivered to the early Earth.

\begin{dmath}\label{minmass2}
dm_{minB}(t) =  0.76\left(1 + \frac{4.57 \times 10^{-7}}{\tau_a} e^{(4.5 - t)/\tau_a} \right) m_2^{1 - b} 4 \pi R_{moon}^2 \Sigma dt
\hspace{3mm}
\textrm{[kg].}
\end{dmath}

\begin{dmath}\label{maxmass2}
dm_{maxB}(t) = 0.4\left(1 + \frac{5.6 \times 10^{-23}}{\tau_c} e^{(4.5 - t)/\tau_c} \right) m_2^{1 - b} 4 \pi R_{moon}^2 \Sigma dt
\hspace{3mm}
\textrm{[kg].}
\end{dmath}
See Figure~\ref{ProbResults} in the main text for a plot of the three mass delivery models.

\subsubsection*{Impactor Mass Distribution}

Chemical analyses of lunar impact samples, and crater size distributions suggest that the impactors of the Earth and Moon before $\sim$3.85 Ga were dominated by main-belt asteroids \cite{2005Sci...309.1847S}. It also likely that the size-frequency distribution of impactors on the early Earth is similar to that of the main-belt asteroids today \cite{2005Icar..175..111B}. Though there is no conclusive evidence to constrain the fraction of early Earth impactors that were of cometary origin, some suggest approximately 10$\%$ of the total accreted mass was from comets \cite{1990Natur.343..129C}. 

The early-Earth impactor mass distribution for impactors with radii 20--40 m, adjusted for the total mass delivered during the LHB, follows the linear relation

\begin{dmath}\label{massdist}
dm_{LHB}(r) = \left[7.5 \times 10^{13} r + 3 \times 10^{15}\right] dr
\hspace{3mm}
\textrm{[kg],}
\end{dmath}
where $dm$ is the mass of impactors with radii $r$ to $r + dr$ (m) \cite{2009Natur.459..419A}.

To get the impactor mass distributions for the mass delivered between $t$ and $t + dt$ in each of our three models, we multiply Equation~\ref{massdist} by Equation~\ref{LHBmass}, \ref{minmass2}, or \ref{maxmass2}, and divide by $2 \times 10^{20}$ kg.

\begin{equation}\label{massdist2}
dm_{i}(t,r) = \left[7.5 \times 10^{13} r + 3 \times 10^{15}\right] \frac{dm_{i}(t)}{2 \times 10^{20}} dr
\hspace{3mm}
\textrm{[kg],}
\end{equation}
where $i$ is the model (LHB, minB, or maxB).

\subsubsection*{Impactor Number Distribution}

Equation~\ref{massdist2} can be turned into a number distribution (from $t$ to $t + dt$) for asteroids of a specific size and type (e.g. carbonaceous chondrites, ordinary chondrites, irons) by multiplying by the fraction of impactors that are asteroids ($f_a$) and the fraction of asteroid impacts that are of a specific meteorite parent body-type ($f_t$), and then dividing by the volume and average density of such asteroids. After simplification, this is

\begin{equation}\label{numdist}
dN_i(t,r) = \frac{3 f_a f_t}{4 \pi \rho} \left[\frac{7.5 \times 10^{13}}{r^2} + \frac{3 \times 10^{15}}{r^3 }\right] \frac{dm_{i}(t)}{2 \times 10^{20}} dr,
\end{equation}
where $i$ is the model (LHB, minB, or maxB).

\subsubsection*{Total Number of Carbonaceous Impactors}

Although some carbonaceous chondrites may have originated from comets \cite{2001PNAS...98.2138E}, for this calculation we conservatively assume that all carbonaceous meteorites originated from asteroids.

Integrating Equation~\ref{numdist} from $r_{min}$ to $r_{max}$ gives the total number of meteoroids in that radii range to have impacted the early Earth, from $t$ to $t + dt$, for each model.

\begin{dmath}\label{intnumdist}
dN_i(t) = \frac{3 f_a f_t}{4 \pi \rho} \left[7.5 \times 10^{13} \left( \frac{1}{r_{min}} - \frac{1}{r_{max}} \right) + \frac{3 \times 10^{15}}{2} \left( \frac{1}{r_{min}^2} - \frac{1}{r_{max}^2} \right) \right] \frac{dm_{i}(t)}{2 \times 10^{20}},
\end{dmath}
where $i$ is the model (LHB, minB, or maxB).

\subsubsection*{Probability of WLP Deposition}

The probability of an infinitesimally small object hitting within a target area, given the probability of hitting anywhere within the total area is equal, is the geometric probability, 

\begin{equation}\label{singprob}
P_g = \frac{A_{targ}}{A_{tot}}.
\end{equation}

This equation can be used to estimate the probability of blindly hitting the bull's eye on a dart board from relatively close in, where $A_{targ}$ is the area of the bull's eye and $A_{tot}$ is the area of the dart board. In this case, since the tip of a dart is relatively small with respect to the bull's eye, the approximation is relatively accurate. In the case of estimating the probability of meteoroid fragments falling into a WLP, the tip of the "dart" (i.e. the debris area of meteoroid fragments) is not always going to be larger than the "bull's eye" (i.e. the total surface area of WLPs at any time). For example, the probability of fragments from a single meteoroid falling into an WLP at $\sim$4.5--4.45 Ga, when there was a small fraction of continental crust and few WLPs, is more analogous to the probability of blindly hitting the bull's eye with a small ball; which would be slightly more likely than with a dart. Any large enough part of the ball overlapping with the bull's eye counts as a hit, as does any large enough part of the debris field overlapping with the combined WLP surface area. ``Large enough'' in our case corresponds to the area of the largest WLP for which the meteoroid fragment deposition probability is being calculated (or equivalently 100 of the smallest WLP in our study). (This minimum area is necessary in order to assume a homogeneous surface deposition when calculating the total mass of meteoroid fragments to have entered a WLP.) 

In order to use Equation~\ref{singprob} to calculate the probability for the fragments of a single meteoroid landing in a primordial WLP, $A_{targ}$ must be the {\it effective} target area. The effective target area is illustrated in Figure~\ref{BullsEye} below. Any meteoroid that enters the atmosphere within the effective target area (of radius $d$), disperses its fragments over at least one WLP's entire surface.

\begin{figure}[ht!]
\centering
\includegraphics[width=\linewidth]{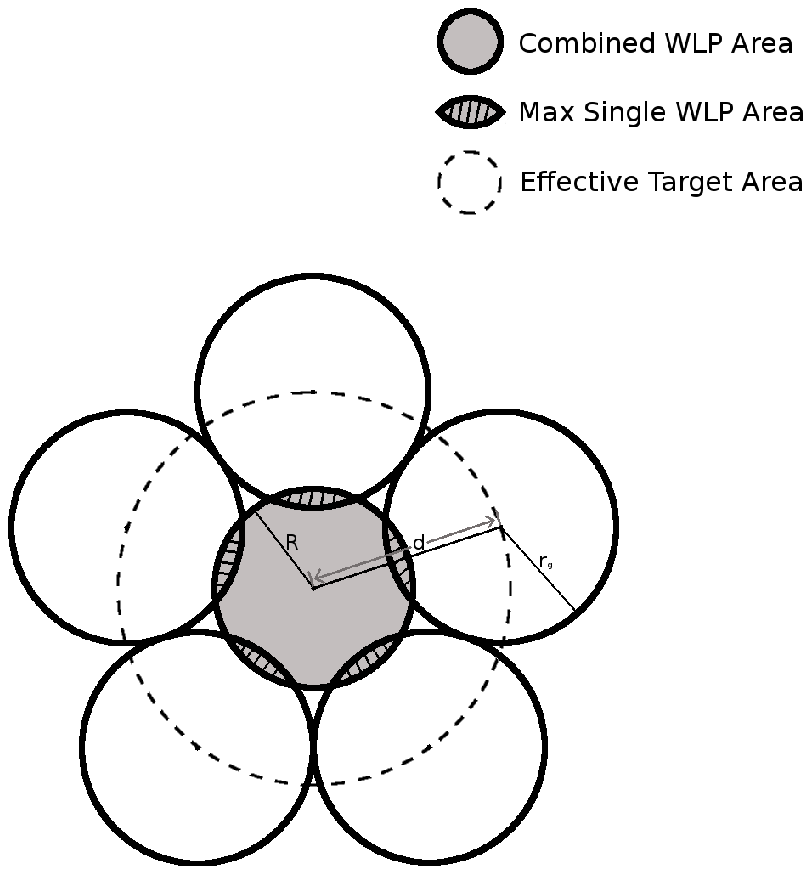}
\caption{The dotted circle represents the target area for landing the dispersed fragments from a single meteoroid into any WLP on the early Earth. The smaller, light grey circle in the center is the total combined WLP surface area on the early Earth at a given time (WLPs would have been individually scattered, however for a geometric probability calculation they can be visualized as a single pond cross-section). The larger, surrounding circles represent the area of debris when the fragments of the meteoroid hit the ground. Shortly after 4.5 Ga, the area of debris from a single meteoroid would be large compared to the the combined WLP surface area. At any time, the target area for a meteoroid to deposit fragments homogeneously into at least 1 WLP is slightly larger than the combined area of WLPs. The effective target area grows linearly with total WLP surface area. The diagonally striped intersections between the circles represent the largest individual WLP for which the meteoroid fragment deposition probability is being calculated. The logic is that if a meteoroid enters the atmosphere at distance $d$ from the center of the combined pond cross-section, at least 1 pond of any size in the WLP distribution will completely overlap with the area of debris.}
\label{BullsEye}
\end{figure}

The area of the asymmetric ``lens'' in which any two circles intersect is

\begin{dmath}\label{areaOfLens}
A = r^2 cos^{-1}\left(\frac{d^2 + r^2 - R^2}{2dr}\right) + R^2 cos^{-1}\left(\frac{d^2 + R^2 - r^2}{2dR}\right) - \frac{1}{2} \sqrt{(-d + r + R) (d + r - R) (d - r + R) (d + r + R)},
\end{dmath}
for circles of radii $r$ and $R$, and distance between their centers $d$ \cite{weisstein2002crc}.

In Figure~\ref{BullsEye}, the asymmetric lens created by the intersection of the combined WLP surface area at a given time and the meteoroid fragment debris area corresponds to the area of the largest individual WLP in our distribution. Because the effective target area radius, $d$, grows linearly with total WLP surface area radius, $R$, we can plot Equation~\ref{areaOfLens} at $t$ = 1 Gyr to solve for $d$, and input the linear time dependence afterwards. For this calculation, $A$ = 3.14 $\times$ 10$^{-4}$ km$^2$, $r$ = $r_g$ = 0.5 km, and $R$ = $\sqrt{\frac{651 km^2}{\pi}}$ = 14.4 km. This gives us the effective target radius, $d$ = 14.9 km, and corresponds to an effective target area:

\begin{equation}\label{Effective}
A_{eff}(t) = 697 t
\tagaddtext{[km$^2$].}
\end{equation}
See Figure~\ref{ProbResults} in the main text for a plot of the effective WLP surface area over time.

Finally, the probability of the fragments from all CM-, CI-, or CR-type meteoroids with radii $r_{min}$ = 20 m to $r_{max}$ = 40 m landing in WLPs on the early Earth of radii 1--10 m from time $t$ to $t + dt$ is the product of Equations~\ref{intnumdist} and \ref{singprob}.

\begin{equation}\label{totalprob}
dP_i(t) = \frac{697 t}{4 \pi R_{\oplus}^2} dN_i(t),
\end{equation}
where $i$ is the model (LHB, minB, or maxB).

\begin{figure}[hbtp]
\centering
\includegraphics[width=\linewidth]{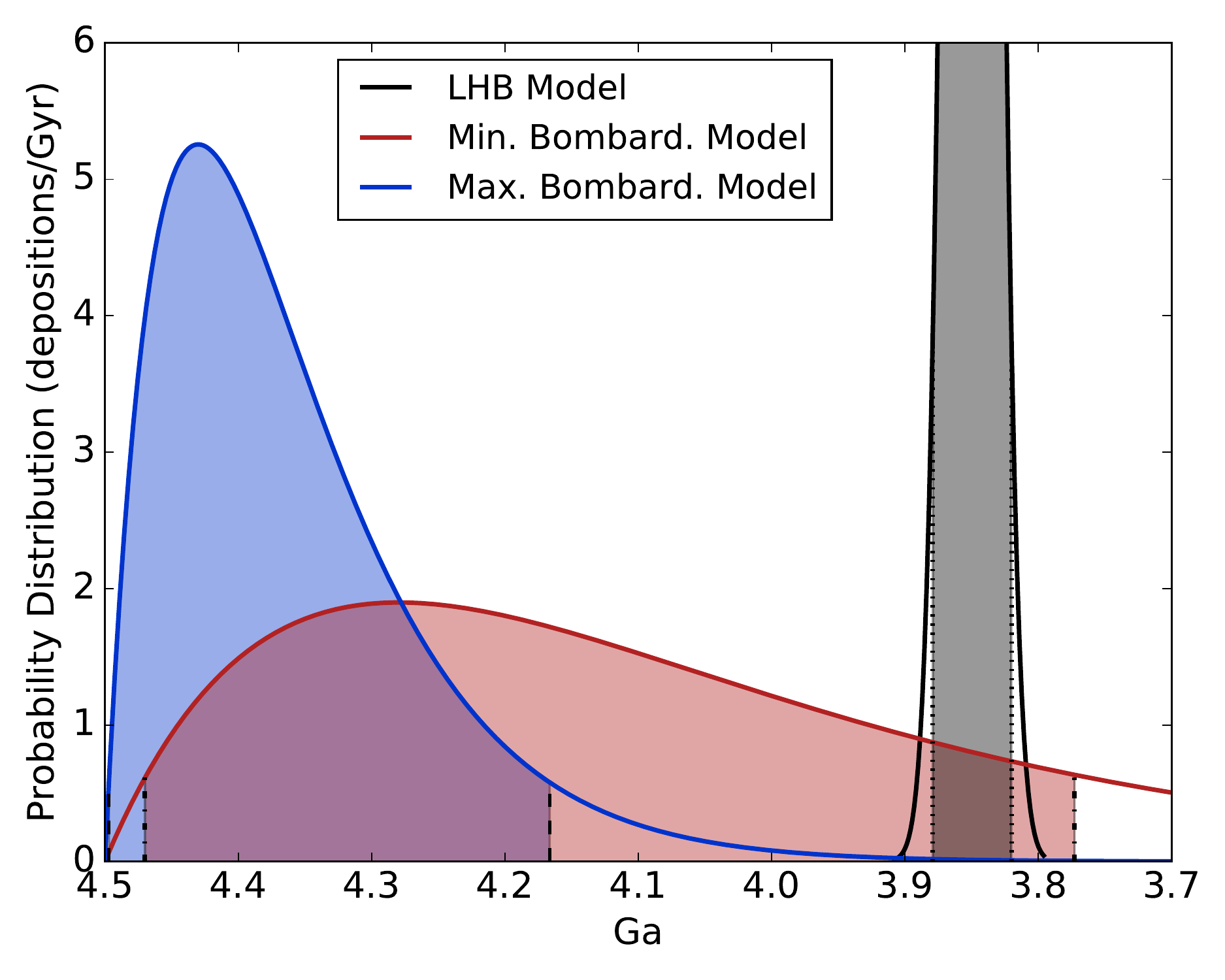}
\caption{Normalized probability distributions of fragments from CM-, CI-, and CR-type meteoroids with radii 20--40 m landing in WLPs on the early Earth of radii 1--10 m. Three models for mass delivery are compared: the Late Heavy Bombardment model, and minimum and maximum mass models for a sustained, declining bombardment preceding 3.9 Ga. All models are based on analyses of the lunar cratering record \cite{2009Natur.459..419A,1990Natur.343..129C}. See Figure~\ref{ProbResults} in the main text for display of mass delivery rates. The 95$\%$ confidence intervals are shaded, and correspond to the most likely deposition intervals for each model.}
\label{ProbDist}
\end{figure}

In Figure~\ref{ProbDist}, we plot the normalized probability distributions ($dP_i/dt$) for the deposition of carbonaceous meteorites into WLPs from 4.5--3.7 Ga. The LHB, the minimum and the maximum bombardment models are compared. For the LHB model, there are 10 WLP depositions from 3.9--3.8 Ga. 95$\%$ of these depositions occur between 3.88--3.82 Ga. For the minimum bombardment model, there are 15 WLP depositions during the entire Hadean eon, however 95$\%$ of these depositions occur between 4.47--3.77 Ga. Finally, for the maximum bombardment model, there are 3840 depositions during the Hadean eon, with 95$\%$ of these depositions occurring between 4.50--4.17 Ga. See Figure~\ref{ProbResults} in the main text for cumulative deposition distributions.

One assumption is made in our deposition probability calculation, which is, given that the largest WLP considered is completely within the carbonaceous meteoroid's strewnfield, at least one fragment will land in the WLP. (An equivalent assumption would be, given that 100 of the smallest WLPs considered are completely within the meteoroid's strewnfield, at least one fragment will deposit into one of the ponds.) To determine whether this is the case, we need a good estimate for the number of fragments that spread over a debris field from a single carbonaceous meteoroid. For the Pu\l{}tusk stony meteoroid, which entered Earth's atmosphere above Poland in 1868 \cite{1971Metic...6..149L}, the number of fragments is estimated to be 180,000 \cite{1971Metic...6..149L}. Although the Pu\l{}tusk meteorites may not represent the typical fragmentation of stony meteoroids, since this meteoroid is about 1.6 times denser than the average carbonaceous meteoroids in our study \cite{2006M&PS...41..331C}, we would expect a carbonaceous meteoroid to fragment more than the typical stony meteorite. If meteoroid fragments are randomly spaced within a strewnfield, then Equation~\ref{singprob} multiplied by the the number of meteoroid fragments will give us roughly the number of fragments that will enter a WLP. (In this case $A_{targ}$ is the area of the cylindrical pond, and $A_{tot}$ is the area of the strewnfield.) From this calculation, given a 10 m-radius WLP is within a 180,000-fragment strewnfield of radius 500 m, roughly 72 carbonaceous meteoroid fragments will land in the pond. In fact, as long as 40--80 m carbonaceous meteoroids fragment into at least 2500 pieces, fragment deposition into a 10 m-radius WLP is probable. We therefore consider this assumption reasonable.

\subsubsection*{Sensitivity Analysis}

In our estimate of the number of carbonaceous meteoroids that led to WLP depositions on the early Earth, we assume that from 4.5--3.7 Ga, the continental crust grows at 16$\%$/Gyr \cite{Reference136}, and the ponds per unit area is the same as today. However, the uncertainty in the growth rate of continental crust and the WLPs per unit area on the early Earth is high. For example, the number of WLPs per unit area was probably higher on the above-sea-level crust of the Hadean Earth than it is on Earth's continents now. The greater rate of asteroid bombardment during the Hadean \cite{2007SSRv..129...35Z} would have created many small craters, which over a few rain cycles could fill to become ponds. On the other hand, geophysical models suggest the surface ocean was increasing in volume from $\sim$4.5--4.0 Ga \cite{Reference92}. This would slow the growth rate of above sea-level crust. Because of these uncertainties, we adjust the growth rate of WLPs on the early Earth by $\pm$ 1 order of magnitude to obtain error bars. Because the probability of deposition (Equation~\ref{totalprob}) is directly proportional to the number of WLPs on the early Earth, varying the growth rate of WLPs by $\pm$ 1 order of magnitude equates to a $\pm$ 1 order of magnitude uncertainty in the WLP deposition expectation values (see Figure~\ref{ProbResults} in the main text).

\subsection*{Sources and Sinks Model Overview}

Calculating the water and nucleobase content in WLPs over time is a problem of sources and sinks. These sources and sinks are illustrated in Figure~\ref{WLPMainFig} in the main text, and displayed in Table~\ref{SourceSink} below.

\begin{table}[!ht]
\centering
\caption{Sources and sinks of pond water and nucleobases in our model of early Earth WLPs. \label{SourceSink}}

\begin{tabular}{ccc}
\\
\multicolumn{1}{c}{Ingredient} &
\multicolumn{1}{c}{Sources} &
\multicolumn{1}{c}{Sinks} \\ \hline \\[-2.5mm]
Pond water & Precipitation & Evaporation \\
 &  & Seepage \\[+2.5mm]
Nucleobases & IDPs & Hydrolysis$^{*}$ \\ 
 & Meteorites$^{*}$ & Seepage$^{*}$ \\
 & & Photodissociation$^{\dagger}$ \\
 & & Forming nucleotides$^{\ddagger}$ \\
\hline
\multicolumn{3}{l}{\footnotesize *Only when pond is wet} \\
\multicolumn{3}{l}{\footnotesize $\dagger$ Only when pond is dry} \\
\multicolumn{3}{l}{\footnotesize $\ddagger$ To be added in a future model}
\end{tabular}
\end{table}

Because nucleobase diffusion from carbonaceous meteorite fragments is slow (see section on nucleobase outflow and mixing below), this source is only turned on in the wet phase. Although it may be possible for nucleobases to occasionally enter WLPs from runoff, we do not consider this as a source. IDPs that fall on dry land are the most likely nucleobase source to be carried by runoff into WLPs (because of their low mass). However, these IDPs are exposed to photodissociating UV light until they are picked up by a runoff stream, by which time few, if any enclosed nucleobases likely remain. Hydrolysis of nucleobases only occurs in the presence of liquid water, therefore we only turn on this sink when our ponds are wet. Similarly, seepage of nucleobases through the pores in the bases of WLPs only occurs during the wet phase. A 1 m column of pond water can absorb UV radiation up to $\sim$95$\%$ \cite{Reference114}, therefore as a first order approximation, we only turn on UV photodissociation when our WLPs are in the dry phase. We do not consider cosmic rays as a sink for nucleobases, as a study has shown that adenine has a half-life of millions of years from cosmic ray dissociation at 1 AU \cite{2011ApJ...730...69E}. This is several orders of magnitude greater than the timescales of nucleobase decay from the other sinks. Finally, since we are interested in accumulating nucleobases in WLPs so that they can react to form nucleotides, a future model will include such reactions as a sink.

\subsection*{Pond Water Sources and Sinks}

\subsubsection*{Evaporation}

There are many variables that could go into a pond evaporation calculation. However, a simple relation was obtained by measuring the depth and temperature of a $\sim$1 m-radius lined pond (to prevent seepage) and a class-A cylindrical evaporation pan over time:

\begin{equation}\label{pondT}
\hspace{23mm}
\frac{dE}{dt} = -9.94 + 5.04T
\hspace{3mm}
\textrm{[mm month$^{-1}$],}
\end{equation}
where $dE$ is the drop in pond depth, and $T$ is the pond temperature in $^{\circ}$C \cite{Reference99}.

This relation is converted to m yr$^{-1}$ below.

\begin{equation}\label{pondT2}
\frac{dE}{dt} = -0.12 + 0.06T
\tagaddtext{[m yr$^{-1}$].}
\end{equation}

\subsubsection*{Seepage}\label{SeepSeep}

Unless the material (e.g. basalt, soil, clay) at the base of a WLP is saturated in water, gravity will cause pond solution to seep through the pores in this material. The average seepage rate of 55 small ponds in Auburn, Alabama was measured to be 5.1 mm day$^{-1}$ \cite{Reference116}. This value is high compared to the average seepage rates from small ponds in North Dakota and Minnesota (1.0 mm day$^{-1}$), and the Black Prairie region of Alabama (1.6 mm day$^{-1}$) \cite{Reference116}. These seepage rates are comparable in magnitude to the pond evaporation rates above, and therefore must be considered as a sink for water and nucleobases in our WLP model. We take an average of the above three values and apply a constant seepage rate of 2.6 mm day$^{-1}$ to our WLP water evolution calculations.

\begin{equation}\label{pondSeep}
\frac{dS}{dt} = 0.95
\tagaddtext{[m yr$^{-1}$].}
\end{equation}
We assume that the majority of seepage occurs from the base of the pond, where the water pressure is the highest. The effect of seepage on nucleobase mass loss is handled in the nucleobase sinks section below.

\subsubsection*{Precipitation}

If we ignore the possibility of local water geysers, the main source of pond water is likely to be precipitation. It has been shown that the vast majority of monthly precipitation climates around the world can be adequately described by a sinusoidal function with a 1 year period \cite{Reference113}. That is,

\begin{equation}\label{precip}
\hspace{12mm}
\frac{dP}{dt} = \overline{P} \left[ 1 + \delta_p sin \left( \frac{2 \pi (t - s_p )}{\tau_s} \right) \right]
\hspace{3mm}
\textrm{[m yr$^{-1}$],}
\end{equation}
where $dP$ is the amount of precipitation, $\overline{P}$ is the mean precipitation rate (m yr$^{-1}$), $\delta_p$ is the dimensionless seasonal precipitation amplitude, $s_p$ is the phase shift of precipitation (yrs), $t$ is the time (yrs), and $\tau_s$ is the duration of the seasonal cycle (i.e. 1 yr) \cite{Reference113}.

From 1980--2009, the mean precipitation on the Earth ranged from $\sim$0.004--10 m yr$^{-1}$ (with a global mean of $\sim$0.7 m yr$^{-1}$), the seasonal precipitation amplitudes ranged from $\sim$0--4.7, and the latitude-dependent phase shift ranged from  0--1 year \cite{Reference113}.

\subsubsection*{Summary}

The evaporation and seepage rates minus the rate of water rise due to precipitation gives us the overall rate of water decrease in a WLP.

\begin{equation}\label{rateOfWaterDrop}
\frac{dL}{dt} = 0.83 + 0.06T - \overline{P} \left[ 1 + \delta_p sin \left( \frac{2 \pi (t - s_p )}{\tau_s} \right) \right]  
\hspace{3mm}
\textrm{[m yr$^{-1}$].}
\end{equation}

Equation~\ref{rateOfWaterDrop} is too complex to be solved by standard analytical techniques or mathematical software, therefore we solve it numerically using a forwards time finite difference approximation with the boundary conditions 0 $\le L(T,t) \le r_p$. We then convert the drop in pond depth as a function of time, $L(t)$, to the mass of water in the WLP using the water density and the volume of a cylindrical portion. This equates to:

\begin{equation}\label{massPond}
m(t) = \pi \rho_w r_p^2 (r_p - L)
\tagaddtext{[kg].}
\end{equation}
For our models we assume pond water of density $\rho_w$ = 1000 kgm$^{-3}$. 

\subsection*{Nucleobase Sinks}

\subsubsection*{Hydrolysis}

The first-order hydrolysis rate constants for adenine (A), guanine (G), uracil (U), and cytosine (C) have been measured from decomposition experiments at pH 7 \cite{Reference45}. These rate constants are expressed in the Arrhenius equations below.

\begin{equation}\label{ade}
k_A = 10^{\frac{-5902}{T} + 8.15} 
\end{equation}
\begin{equation}
k_G = 10^{\frac{-6330}{T} + 9.40}
\end{equation}
\begin{equation}
k_U = 10^{\frac{-7649}{T} + 11.76}
\end{equation}
\begin{equation}\label{cyt}
k_C = 10^{\frac{-5620}{T} + 8.69}
\tagaddtext{[s$^{-1}$].}
\end{equation}

The nucleobase decomposition rate due to being dissolved in water can then be obtained by plugging Equations~\ref{ade}--\ref{cyt} into the first-order reaction rate law below.

\begin{equation}\label{ratelaw}
\frac{dm_i}{dt} = -m_i \gamma k 
\tagaddtext{[kg yr$^{-1}$],}
\end{equation}
where $\gamma$ = 3600 $\cdot$ 24 $\cdot$ 365.25 s yr$^{-1}$.

The hydrolysis rates of adenine, guanine, and cytosine remain relatively stable in solutions with pH values from 4.5--9 \cite{Reference45,2011OLEB...41..553K}. WLPs may have been slightly acidic (from pH 4.8--6.5) due to the higher partial pressure of CO$_2$ in the early Earth atmosphere \cite{Reference80}.

\subsubsection*{UV photodissociation}

The photodestruction of adenine has been studied by irradiating dried samples under Martian surface UV conditions \cite{2015AsBio..15..221P}. This quantum efficiency of photodecomposition (from 200--250 nm)---which is independent of the thickness of the sample---was measured to be $1.0 \pm 0.9 \times 10^{-4}$ molecule photon$^{-1}$ \cite{2015AsBio..15..221P}.

For the calculation of the quantum efficiency of adenine, a beam of UV radiation was focused on a thin compact adenine sample formed through sublimation and recondensation \cite{2015AsBio..15..221P}. In this case, all the photons in the beam of UV radiation were incident on the nucleobases. In the WLP scenario, not all the incoming UV photons will be incident on nucleobases. Instead, large gaps can exist between nucleobases that collect at the base of the pond. Therefore the number of photons incident on the pond area is not the same as those incident on the scattered nucleobases unless there are at least enough nucleobases present to cover the entire pond area. Since the nucleobases are mixed well into the pond water before complete evaporation, we assume nucleobases will spread out evenly as they collect on the base of the pond. This means we assume that all locations on the base of the pond are covered in nucleobases before nucleobases stack on top of one another. For low abundances of total nucleobases in a WLP, this approach will lead to a slightly higher estimate for the rate of photodissociation than expected. We deem this acceptable for a first-order estimate of nucleobase photodissociation.

The mass of nucleobases photodestroyed per year, per area covered by nucleobases (i.e. the photodestruction flux), is constant over time and is dependent on the experimentally measured quantum efficiency of photodecomposition.

\begin{dmath}
\dot M_i = 
      \cfrac{\Phi F \lambda \gamma \mu_i}{h c N_A}
     \hspace{3mm}
\textrm{[kg yr$^{-1}$ m$^{-2}$],}
\end{dmath}
where $\Phi$ is the quantum efficiency of photodecomposition of the molecules (molecules photon$^{-1}$), $F$ is the UV flux incident on the entire pond area (W m$^{-2}$), $\lambda$ is the average wavelength of UV radiation incident on the sample (m), $\gamma$ is 3600 $\cdot$ 24 $\cdot$ 365.25 s yr$^{-1}$, $\mu_i$ is the molecular weight of the irradiated molecules (kg mol$^{-1}$), $h$ is Planck's constant (m$^2$ kg s$^{-1}$), $c$ is the speed of light (ms$^{-1}$), and $N_A$ is Avogadro's number (molecules mol$^{-1}$).

Our estimation of the photodestruction rate of a nucleobase depends on the total mass of the nucleobase within the WLP. If there is enough of the nucleobase present for the entire base of the pond to be covered, we can multiply the photodestruction flux by the entire pond area. Otherwise, we must multiply the photodestruction flux by the combined cross-sectional area of the nucleobase present in the WLP to get the photodestruction rate.

\begin{equation}
\cfrac{dm_i}{dt} = 
\begin{cases}
     - \dot M_i \cfrac{m_i}{\rho_i d}, & \text{if}\ \cfrac{m_i}{\rho_i d} < A_p \\
     - \dot M_i A_p, & \text{otherwise}
\end{cases}
\hspace{3mm}
\textrm{[kg yr$^{-1}$],}
\end{equation}
where $m_i$ is the mass of the sample (kg), $\rho_i$ is the mass density of the nucleobase (kg m$^{-3}$), $d$ is the distance between two stacked nucleobases in the solid phase (m), and $A_p$ is the area of the WLP (m$^2$).

Assuming cloudless skies, an upper limit on the early Earth integrated UV flux from 200--250 nm is $\sim$0.4 W m$^{-2}$ \cite{2015ApJ...806..137R,2002abqc.book..219C}. UV wavelengths $<$ 200 nm would be completed attenuated by CO$_2$ and H$_2$O in the early atmosphere \cite{2015ApJ...806..137R}. The mass density of solid adenine is 1470 kg m$^{-3}$, making the distance between two stacked adenine molecules in the solid phase $\sim$6.6 $\AA$. For guanine, uracil, and cytosine, we take the mass densities to be 2200 kg m$^{-3}$, 1320 kg m$^{-3}$, and 1550 kg m$^{-3}$, respectively.

\subsubsection*{Seepage}\label{nucSeep}

The constant pond water seepage $\dot S$ = 0.95 m yr$^{-1}$ was determined in the pond water source and sinks section above. This can be used to calculate the nucleobase seepage rate via the equation:

\begin{equation}
\cfrac{dm_i}{dt} =  w_i \rho_w A_p \dot S 
\tagaddtext{[kg yr$^{-1}$],}
\end{equation}
where $w_i$ is the nucleobase mass fraction, $\rho_w$ is the density of water (kg yr$^{-1}$), and $A_p$ is the area of the WLP (m$^2$).

\subsection*{Nucleobase Outflow and Mixing}\label{diffusion}

Chondritic IDPs and meteorites are porous \cite{1997M&PS...32..509C,2006M&PS...41..903M,2009E&PSL.287..559B}. With the exception of the nucleobases potentially formed due to surface photochemistry \cite{2014ApJ...793..125N}, any soluble nucleobases delivered to the prebiotic Earth by carbonaceous IDPs and meteorites would have layed frozen in the pores of these sources upon them entering the atmosphere. Pulse heating experiments show approximately $\sim$1--6$\%$ of the organics within IDPs could have survived atmospheric entry \cite{2006M&PS...41..903M}. And, based on their presence in carbonaceous chondrites today, nucleobases in carbonaceous meteorites evidently would have also survived atmospheric entry heating. Therefore both sources, upon entering WLPs on the prebiotic Earth, would have slowly released their remaining soluble nucleobases into the surrounding pond water. These nucleobases would then slowly homogenize into the pond solution.


We model the outflow of nucleobases from carbonaceous IDPs and meteorites and the mixing of a local concentration of nucleobases into a WLP using finite difference approximations of the one-dimensional advection-diffusion equation (see Appendix A for complete details). 

Our model of nucleobase outflow is run for average-sized IDPs (r = 100$\mu$m), and for small (r = 1 cm), medium (r = 5 cm) and large (r = 10 cm) carbonaceous meteorites. 
The fraction of nucleobases remaining in each of these sources as a function of time is plotted in Figure~\ref{IDPMetDiff} below.

\begin{figure*}[htbp]
\centering
\includegraphics[width=17.8cm]{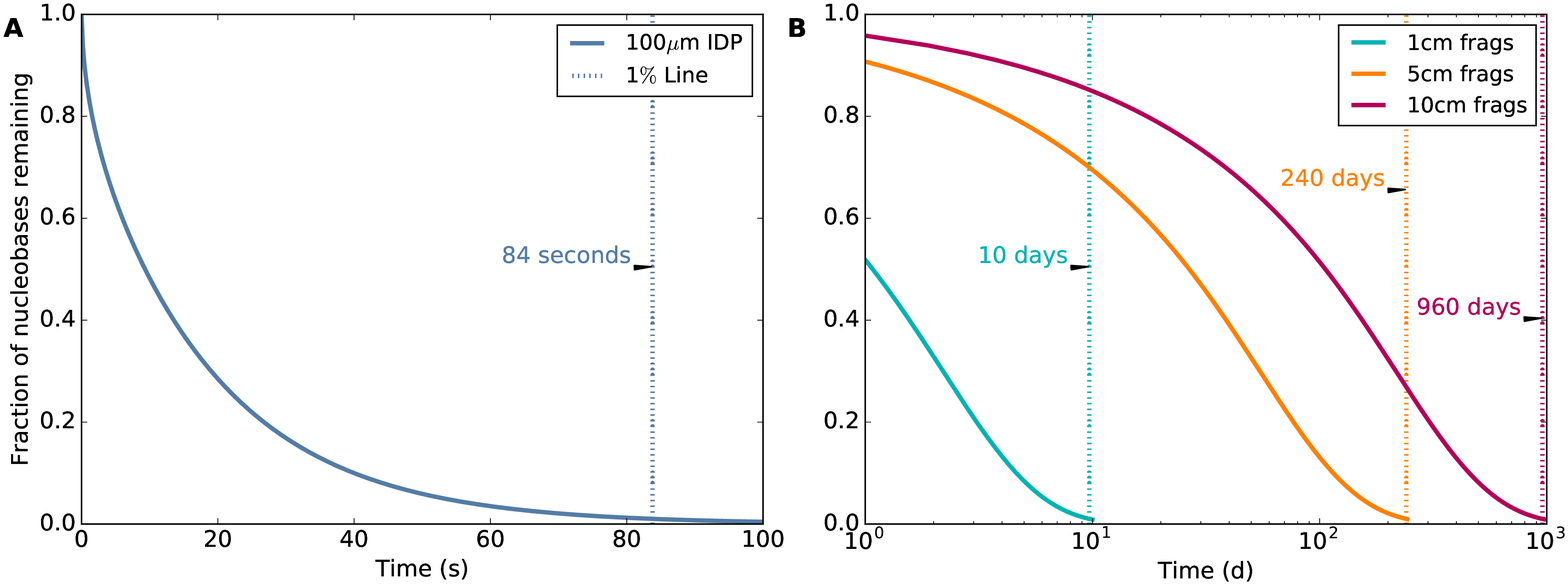}
\caption{Fraction of the total initial nucleobases remaining in {\bf (A)} a 100$\mu$m-radius IDP and {\bf (B)} 1 cm-, 5 cm-, and 10 cm-radii meteorites over time as a result of diffusion across a rock-pond boundary. The IDP and meteorites are considered to be laying on the bottom of a WLP, and are diffusing nucleobases symmetrically in the radial direction. The times at which 99$\%$ of the initial contained nucleobases have diffused into the WLP are labeled on the plots.}
\label{IDPMetDiff}
\end{figure*}

Our models show that the duration of nucleobase diffusion from carbonaceous IDPs and meteorites into WLPs is mostly determined by the radius of the source. For typical, 100 $\mu$m-radius carbonaceous IDPs laying at the bottom of a WLP, it takes $<$ 2 minutes for $>$ 99$\%$ of the soluble nucleobases to diffuse into the surrounding pond water. For 1 cm-radius carbonaceous meteorites, this time increases to 10 days. For the largest carbonaceous meteoroid fragments, with radii of 5 cm and 10 cm, this duration increases to approximately 8 and 32 months, respectively.

\begin{figure}[ht!]
\centering
\includegraphics[width=\linewidth]{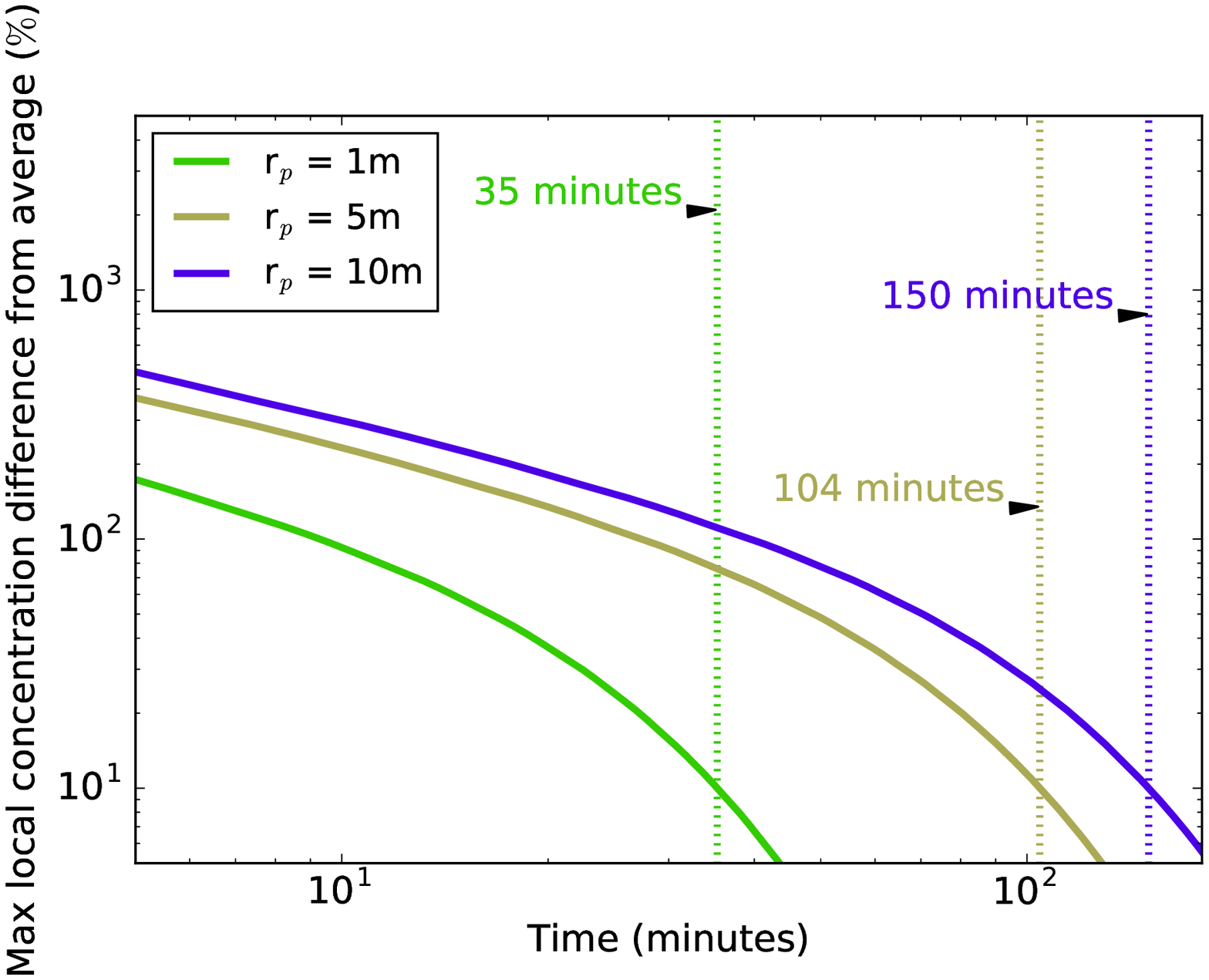}
\caption{The nucleobase mixing time in a base-to-surface convection cell (length = 2$r_p$) within 1 m-, 5 m-, and 10 m-deep WLPs, beginning from a local concentration at the base of the pond. Nucleobase mixing is measured using the maximum percent local nucleobase concentration difference from the average. The time at which the maximum local nucleobase concentration difference from the average drops to 10$\%$ is labeled on the plot for each pond size. At this time, we consider the nucleobases in the WLP to be well mixed. For WLPs with radii 1 m, 5 m, and 10 m, the convection cell nucleobase mixing times are 35 minutes, 104 minutes, and 150 minutes, respectively.}
\label{WLPMix}
\end{figure}

For nucleobase mixing, we model a base-to-surface convection cell within cylindrical ponds with equal radii and depths of 1 m, 5 m, and 10 m. This model gives us the timescale of mixing a local concentration of nucleobases within WLPs. The maximum percent local nucleobase concentration difference from the average is plotted as a function of time in Figure~\ref{WLPMix}. This metric allows us to characterize the nucleobase homogeneity in a convection cell of the WLP.

Our simulations suggest that the mixing of local deposits of nucleobases in WLPs, resulting from their diffusion out of carbonaceous IDPs and meteorites, is a very efficient process. For a cylindrical WLP 1 m in depth, it will take about 35 minutes for a local deposition to homogenize in a convection cell within the pond. For larger WLPs, with 5 m and 10 m depths, the nucleobase mixing time increases to 104 and 150 minutes, respectively. These short mixing times make it clear that, for carbonaceous meteorites $\geq$ 1 cm in radius, nucleobase homogenization in WLPs is dominated by the nucleobase outflow time from these bodies. In contrast, being that carbonaceous IDPs are much smaller than meteoroid fragments, the nucleobase outflow time from IDPs is negligible compared to the nucleobase homogenization time in WLPs.

\subsection*{Nucleobase Evolution Equation From IDPs}\label{IDPTheory}

It is estimated that at 4 Ga, approximately $6 \times 10^8$ kgyr$^{-1}$ of carbonaceous IDPs were being accreted onto the Earth \cite{1992Natur.355..125C}. Since IDPs are tiny (typically $\sim$100 $\mu$m in radius), they are circulated by the atmosphere upon accretion and thus likely reached almost everywhere on the prebiotic Earth. Approximately 1--6$\%$ of the organic content within IDPs could have survived the pulse heating of atmospheric entry \cite{2006M&PS...41..903M}. Assuming IDPs accreted uniformly, the surface nucleobase mass accretion per square area would be

\begin{dmath}\label{accretion}
\frac{dm_i}{dtdA} = \frac{w_i \dot m_{I} f_s}{4 \pi R_{\oplus}^2} 
     \hspace{3mm}
\textrm{[kg yr$^{-1}$m$^{-2}$],}
\end{dmath}
where $w_i$ is the nucleobase mass fraction within IDPs for nucleobase $i$, $\dot m_{I}$ is the mass accretion rate of IDPs on the prebiotic Earth (kg yr$^{-1}$), and $f_s$ is the average fraction of nucleobases that survive pulse heating from atmospheric entry. 

IDPs are thought to correspond to origins of asteroids or comets \cite{1992Natur.355..125C}, therefore at best, the average nucleobase abundances in IDPs could match the average nucleobase content within the nucleobase-rich CM, CR, and CI meteorites. The average abundances of adenine, guanine, and uracil in CM, CR, and CI meteorites are listed in Table~\ref{AvgNucleoAbun} along with the weighted averages based on relative fall frequencies. These abundances might be an upper limit for the guanine, adenine, and uracil content of IDPs because unlike the interior of large meteorites, molecule-dissociating UV radiation can penetrate everywhere within $\mu$m--mm-sized pieces of dust.

\begin{table}[!ht]
\centering
\caption{Average guanine, adenine and uracil abundances (in ppb) in the CM, CR, and CI carbonaceous chondrites. Abundances obtained from \cite{Reference46}. Some CM and CR meteorite analyses found no adenine or uracil, these samples were excluded from the average. Weighted nucleobase averages are also displayed based on relative fall frequencies \cite{2002aste.book..653B}. Uracil has not been measured in CR meteorites. Cytosine has not been measured in any meteorites. \label{AvgNucleoAbun}}

\begin{tabular}{cccc}
\\
\multicolumn{1}{c}{Meteorite Type} &
\multicolumn{1}{c}{Guanine} &
\multicolumn{1}{c}{Adenine} &
\multicolumn{1}{c}{Uracil}\\ \hline \\[-2.5mm]
CM & 183.5 & 69.8 & 50.0 \\
CR & 1.8 & 9.3 & - \\
CI & 81.5 & 60.5 & 73.0 \\[2mm]
Weighted Avg & 141.3 & 60.7 & 48.6\\
\hline
\end{tabular}
\end{table}

The amount of cytosine that could have formed on the surfaces of primordial IDPs is not well constrained. Cytosine has been detected in experiments exposing IDP analogs to UV radiation, but hasn't been quantified \cite{2014ApJ...793..125N}. Furthermore, these analog experiments formed cytosine via photoreactions involving pyrimidine: a molecule that has no measured abundance in IDPs or meteorites. For this analysis we explore a best-case scenario, and set the maximum cytosine IDP abundance to 141.3 ppb (the possible upper limit of guanine in carbonaceous IDPs).

In the nucleobase outflow and mixing section above we learned that nucleobase diffusion from IDPs is quick (lasting $<$ 2 minutes), and nucleobase homogenization in WLPs takes one to a few hours, depending on the pond size. If we spread out the $6 \times 10^8$ kgyr$^{-1}$ accretion rate of IDPs uniformly across the surface of the Earth---assuming IDPs are all 100 $\mu$m-radius spheres of CI, CM, and CR type ($\rho$ = $\sim$2185 kgm$^{-3}$)---then IDPs would drop into 1--10 m-radius primordial WLPs at approximately 0.05--5 per hour. Since each carbonaceous IDP can only carry a tiny mass in nucleobases (~$\sim$1 picogram), nucleobase inhomogeneities within WLPs would be a maximum of $\sim$15 picograms. This abundance of nucleobases is negligible, therefore for our calculations of nucleobase accumulation in WLPs from IDP sources, we can assume that nucleobase deposition and pond homogenization is instantaneous.

Thus, the differential equation for the mass of nucleobase $i$ accumulated in a WLP from IDP sources over time is the sum of the nucleobase mass accretion rate from IDPs, the nucleobase mass decomposition rate, the nucleobase mass seepage rate, and the nucleobase photodissociation rate:

\begin{dmath}\label{IDPAccumEq}
\frac{dm_{i,IDP}(t)}{dt} = \frac{w_i \dot m_{I} f_s A_{p}}{4 \pi R_{\oplus}^2} - \gamma m_i k_i - w_i \rho_w A_p \dot S 
- \begin{cases}
      \dot M_i \cfrac{m_i}{\rho_i d}, & \text{if}\ \cfrac{m_i}{\rho_i d} < A_p \\
      \dot M_i A_p, & \text{otherwise}
\end{cases} 
\hspace{3mm}
\textrm{[kg yr$^{-1}$]}.
\end{dmath}
The second and third terms after the equal sign are only activated when the pond is wet, and the fourth term is only activated when the pond is dry. The first term is always activated as recently accreted nucleobases are susceptible to UV dissociation while still laying in the pores of IDPs, and they effectively instantaneously outflow from IDPs upon wetting.

Using Equation~\ref{IDPAccumEq}, we compute the nucleobase mass as a function of time numerically using a forwards time finite difference approximation. We then divide the nucleobase mass by the water mass at each time step to obtain the nucleobase mass concentration over time. Because some ponds are seasonally dry, we freeze the water level at 1 mm during the dry phase in order to calculate a nucleobase concentration during this phase.

\subsection*{Nucleobase Evolution Equation From Meteorites}\label{MetTheory}

Simulations show the fragments from carbonaceous meteoroids with diameters from 40--80 m and initial velocities of 15 km/s will expand over a radius of $\sim$500 m, and $\sim$ 32$\%$ of the original meteoroids will survive ablation \cite{1993AJ....105.1114H}. Since a single meteoroid impacting the atmosphere may break up into 180,000 fragments before spreading across its strewnfield, it is probable that a meteorite deposition event involves multiple meteorites landing in a single WLP. The best estimate of the total nucleobase mass deposited into a WLP is thus calculated assuming the ablated meteoroid mass is spread uniformly over its strewnfield. Considering this, the total nucleobase mass to deposit into a WLP from a meteoroid would be 

\begin{equation}\label{metMass}
m_{i0} = \frac{4}{3} \frac{w_i f_s r^3 \rho A_p}{r_g^2} 
\tagaddtext{[kg],}
\end{equation}
where $w_i$ is the nucleobase mass fraction within the meteoroid for nucleobase $i$, $f_s$ is the fraction of the meteoroid to survive ablation, $r$ is the meteoroid radius as it enters Earth's atmosphere (m), $\rho$ is the density of the meteoroid (kg m$^{-3}$), $r_g$ is the radius of the debris when the meteoroid fragments hit the ground (m), and $A_p$ is the area of the WLP (m$^2$).

After the deposition of meteoroid fragments into a WLP, the frozen meteorite interiors will thaw to pond temperature, allowing hydrolysis to begin inside the fragments' pores. This means the total mass of nucleobases that diffuse from the fragments' pores into the pond will be less than the total initial nucleobase mass within the deposited fragments. By integrating the nucleobase hydrolysis rate (Equation~\ref{ratelaw}), we obtain the mass of nucleobase $i$ remaining after a given time of hydration $t_h$,

\begin{equation}\label{massleft1}
m_i = m_{i0} e^{-\gamma k_i t_h} 
\tagaddtext{[kg].}
\end{equation}
Note that $t_h$ (yr) may be different than $t$, as the hydration clock is paused when WLPs are dry.

Unlike carbonaceous IDPs, which unload their nucleobases into a WLP in seconds, nucleobases may not completely outflow from all deposited carbonaceous meteoroid fragments before the WLP evaporates. (However, the nucleobases that do outflow from the meteorites will mix homogeneously into the WLP within a single day-night cycle.) Thus, we calculate the nucleobase outflow time constants for 1 cm-, 5 cm-, and 10 cm-radius carbonaceous meteorites by performing least-squares regressions of our nucleobase diffusion simulation results to the function below.

\begin{equation}
f(t_h) = \alpha \left(1 - e^{-\frac{t_h}{\tau_d}} \right).
\end{equation}
This equation shows that as time increases, the mass of nucleobases that have flowed out of a single meteorite, into the WLP, approaches the coefficient $\alpha$---which represents the total initial nucleobase mass contained in the meteorite. Since nucleobase outflow is mass-independent, we can use an arbitrary initial total nucleobase mass for $\alpha$ in our simulations to obtain the nucleobase outflow time constants.

The results of the fits give diffusion time constants for 1 cm-, 5 cm-, and 10 cm-radius fragments of $\tau_d$ = 4.9 $\times$10$^{-3}$ yr, 0.12 yr and 0.48 yr, respectively.

Adding up the sources and sinks gives us the nucleobase mass within the WLP from meteorite sources as a function of time and hydration time.

\begin{dmath}\label{MetAccumEq}
\frac{dm_{i,Met}(t, t_h)}{dt} = \frac{m_{i0}}{\tau_d} e^{-t_h \left(\gamma k_i +\frac{1}{\tau_d} \right)} -  \gamma m_i k_i - w_i \rho_w A_p \dot S 
- \begin{cases}
      \dot M_i \cfrac{m_i}{\rho_i d}, & \text{if}\ \cfrac{m_i}{\rho_i d} < A_p \\
      \dot M_i A_p, & \text{otherwise}
\end{cases} 
\hspace{3mm}
\textrm{[kg yr$^{-1}$]}.
\end{dmath}
The first three terms after the equal sign are only activated when the pond is wet, and the fourth term is only activated when the pond is dry.

Since there are many possibilities for the sizes of meteoroid fragments that will enter a WLP, we consider 3 simplified models: all fragments that enter a WLP from a meteoroid of radius 20--40 m are either 1 cm in radius, 5 cm in radius, or 10 cm in radius. These three models represent a local part of the strewnfield that deposited either many small fragments, mostly medium-sized fragments, or just a couple to a few large fragments.

Cytosine is unlikely to have sustained within meteorite parent bodies long enough to be delivered to the early Earth by meteorites \cite{Reference106}, therefore we only model the accumulation of adenine, guanine, and uracil in WLPs from meteorite sources.

We solve Equation~\ref{MetAccumEq} numerically using a forwards time finite difference approximation. Nucleobase concentration is then obtained by dividing the nucleobase mass by the water mass in the WLP at each time step. 

\subsection*{Additional results}

\subsubsection*{Pond Water}

\begin{figure}[ht!]
\centering
\includegraphics[width=\linewidth]{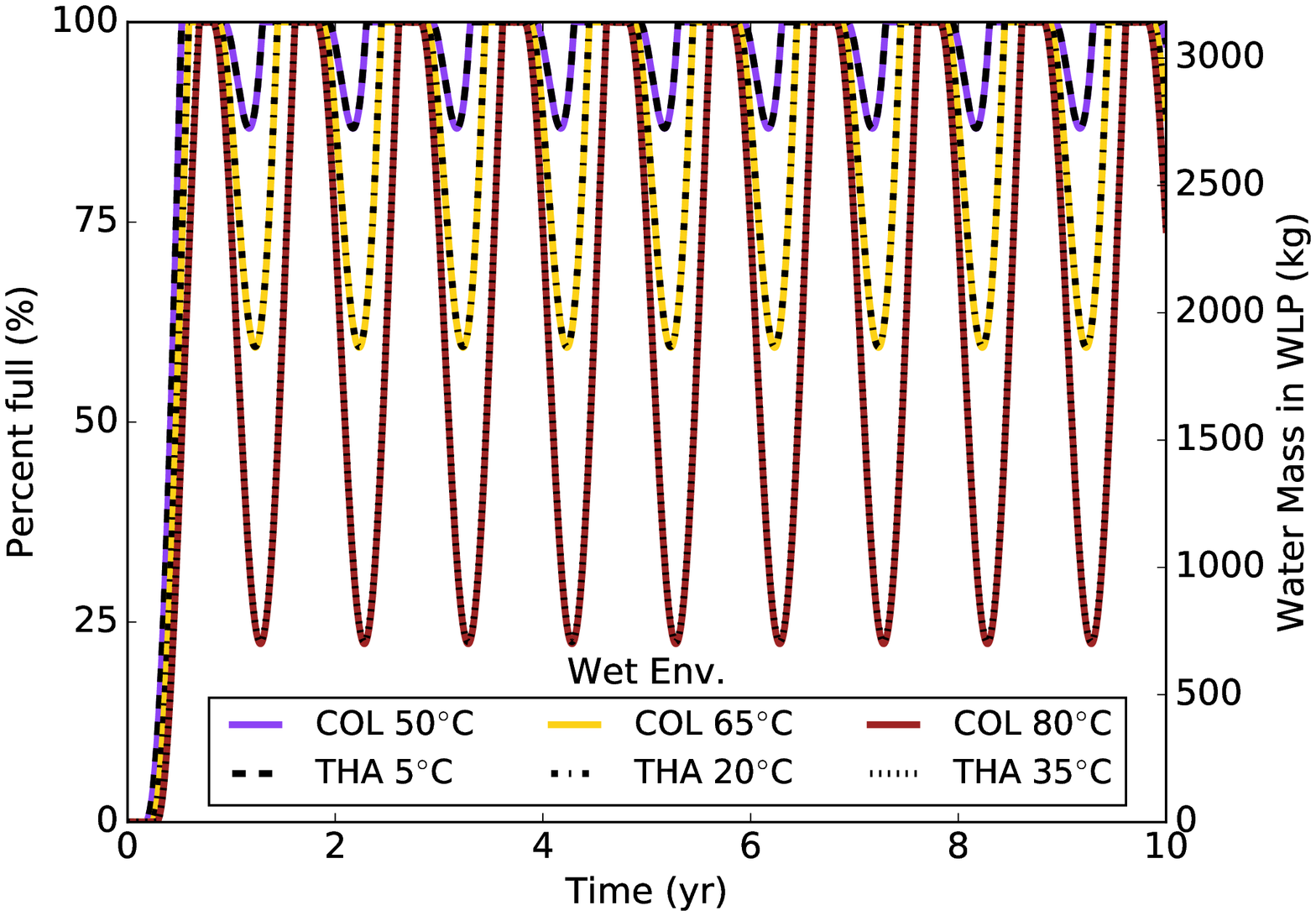}
\caption{The effect of temperature on the change in water mass over time in wet environment cylindrical WLPs with radii and depths of 1 m. Temperatures are varied for a hot early Earth model (50 $^{\circ}$C, 65 $^{\circ}$C, and 80 $^{\circ}$C) and a warm early Earth model (5 $^{\circ}$C, 20 $^{\circ}$C, and 35 $^{\circ}$C). Precipitation rates from Columbia and Thailand on Earth today are used to represent the hot, and warm early Earth analogues, respectively (for details see Table~\ref{precipModels} in the main text).}
\label{WLPWaterTemp}
\end{figure}

In Figure~\ref{WLPWaterTemp} we explore the effects of changing temperature on wet environment WLPs of 1 m radius and depth (see Table~\ref{precipModels} in main text for wet environment model details). To do this, we vary our fiducial model temperatures (65$^{\circ}$C for a hot early Earth and 20$^{\circ}$C for a warm early Earth) by $\pm$15$^{\circ}$C. As temperature increases, evaporation becomes more efficient, however the wet environment WLPs never dry completely.

\subsubsection*{Nucleobase Accumulation from IDPs}

\begin{figure*}[htbp]
\centering
\includegraphics[width=17.8cm]{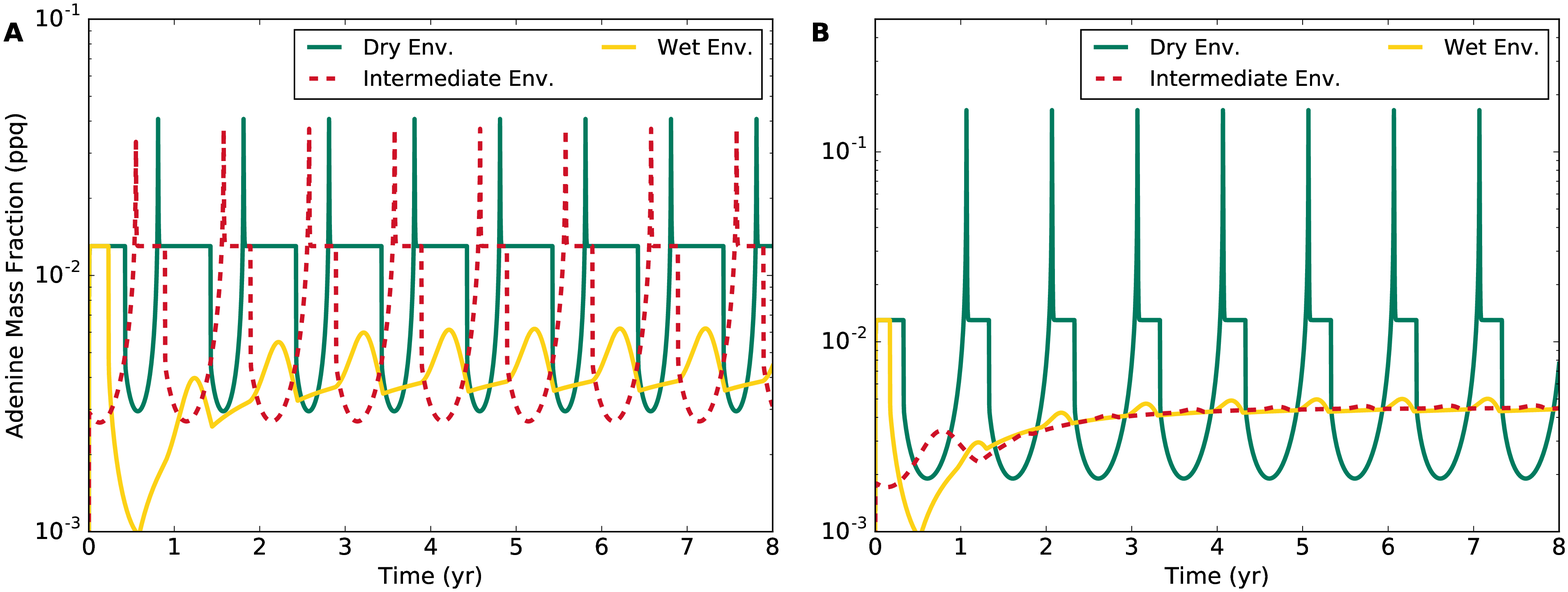}
\caption{The accumulation of adenine from only carbonaceous IDP sources in cylindrical WLPs with radii and depths of 1 m. The three curves (dry, intermediate, and wet environments) differ by their precipitation rates, which are from a variety of locations on Earth today, and represent 2 classes of matching early Earth analogues: hot (Columbia, Indonesia, Cameroon), and warm (Thailand, Brazil, and Mexico) (for details see Table~\ref{precipModels} in the main text). {\bf (A)} The degenerate WLP models used for these calculations correspond to a hot early Earth at 65 $^{\circ}$C and a warm early Earth at 20 $^{\circ}$C. {\bf (B)} The degenerate WLP models used for these calculations correspond to a hot early Earth at 50 $^{\circ}$C and a warm early Earth at 5 $^{\circ}$C.}
\label{IDPAccum}
\end{figure*}

In Figure~\ref{IDPAccum} we explore the evolution of adenine concentration from only IDP sources in WLPs of 1 m radius and depth. We model the adenine accumulation in three environments (dry, intermediate, and wet) on a hot and warm early Earth (see Table~\ref{precipModels} in the main text for precipitation model details). The left panel is for 65$^{\circ}$C on a hot early Earth and 20$^{\circ}$C on a warm early Earth, and the left panel is for 50$^{\circ}$C on a hot early Earth and 5$^{\circ}$C on a warm early Earth.

All adenine concentration curves reach a stable seasonal pattern within $\sim$5 years. The highest adenine concentrations occur for models with a dry phase (i.e. the dry and intermediate models at 65$^{\circ}$C, and the dry model at 50$^{\circ}$C.) The lower maximum concentrations in the models without a dry phase are due to the sustained high water levels. The maximum adenine to accumulate in any model from IDP sources is $\sim$0.2 ppq. This occurs just before the pond dries. Upon drying, UV photodissociation immediately drops the adenine concentration to an amount which balances the incoming adenine from IDP accretion. The curves in Figure~\ref{IDPAccum} do not change drastically with increasing pond radius and depth once a stable seasonal pattern is reached. This is because although ponds of increasing surface area collect more nucleobases, these ponds have an equivalent increase in area to nucleobase seepage.

\begin{figure}[ht!]
\centering
\includegraphics[width=\linewidth]{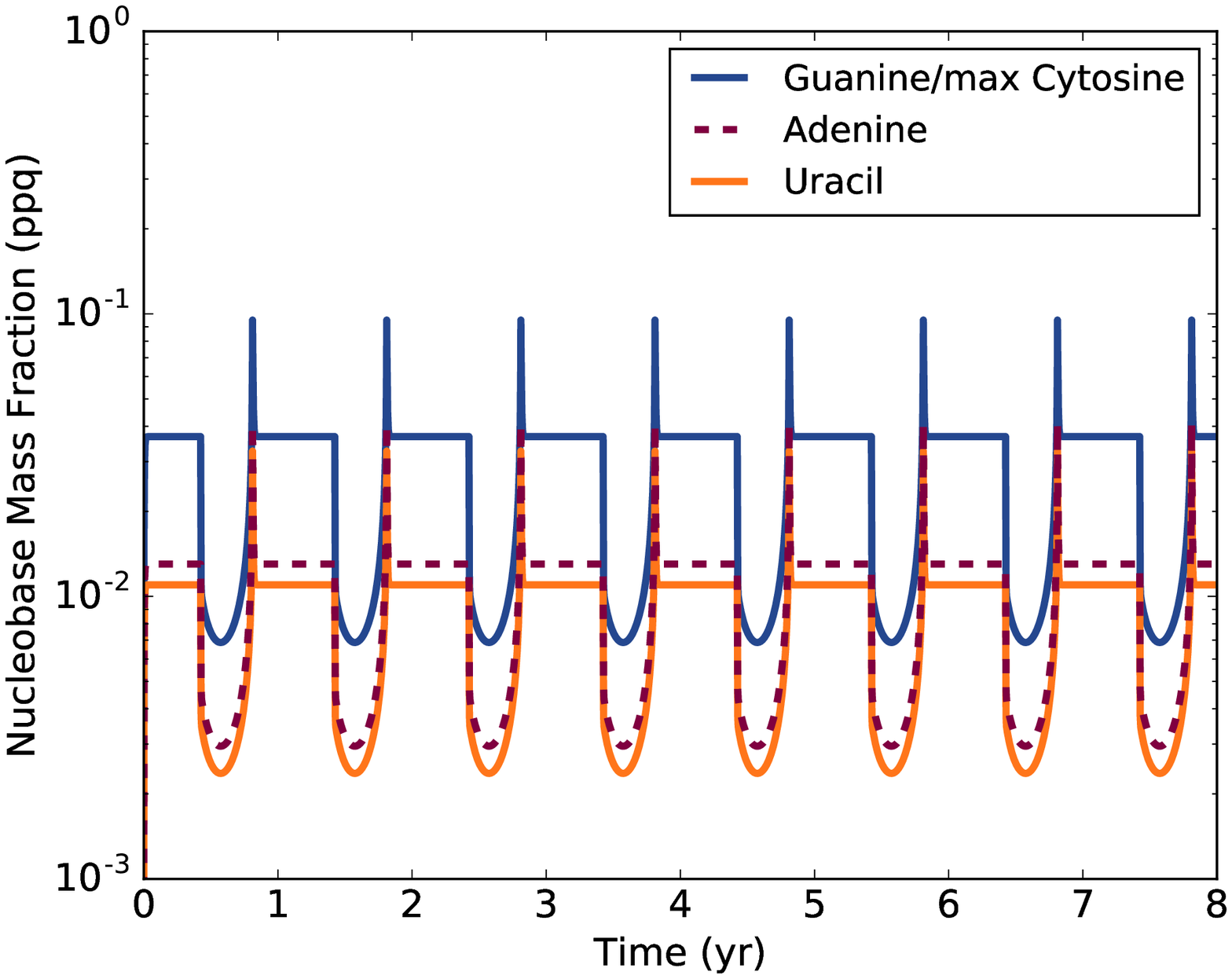}
\caption{The accumulation of guanine, adenine, uracil, and cytosine from only carbonaceous IDP sources in cylindrical WLPs with radii and depths of 1 m. The degenerate dry WLP models used for these calculations correspond to a hot early Earth at 65 $^{\circ}$C and a warm early Earth at 20 $^{\circ}$C. Precipitation rates from Cameroon and Mexico on Earth today are used to represent the hot, and warm early Earth analogues, respectively (for details see Table~\ref{precipModels} in the main text). The cytosine abundance in IDPs used to calculate the max cytosine curve, matches the average abundance of guanine in IDPs (see Table~\ref{AvgNucleoAbun}). The curves are obtained by numerically solving Equation~\ref{IDPAccumEq}.}
\label{IDPAccum65All}
\end{figure}

In Figure~\ref{IDPAccum65All} we explore guanine, uracil, and cytosine accumulation in degenerate dry environment WLPs for a hot early Earth at 65 $^{\circ}$C and a warm early Earth at 20 $^{\circ}$C (See Table~\ref{precipModels} in main text for details). The differences in each nucleobase mass fraction over time is caused by the different initial abundances of each nucleobase in IDPs (see Table~\ref{AvgNucleoAbun}). Although hydrolysis rates differ between nucleobases, decay due to hydrolysis is negligible over $<$ 10 year periods at temperatures lower than $\sim$70 $^{\circ}$C.

\begin{figure}[ht!]
\centering
\includegraphics[width=\linewidth]{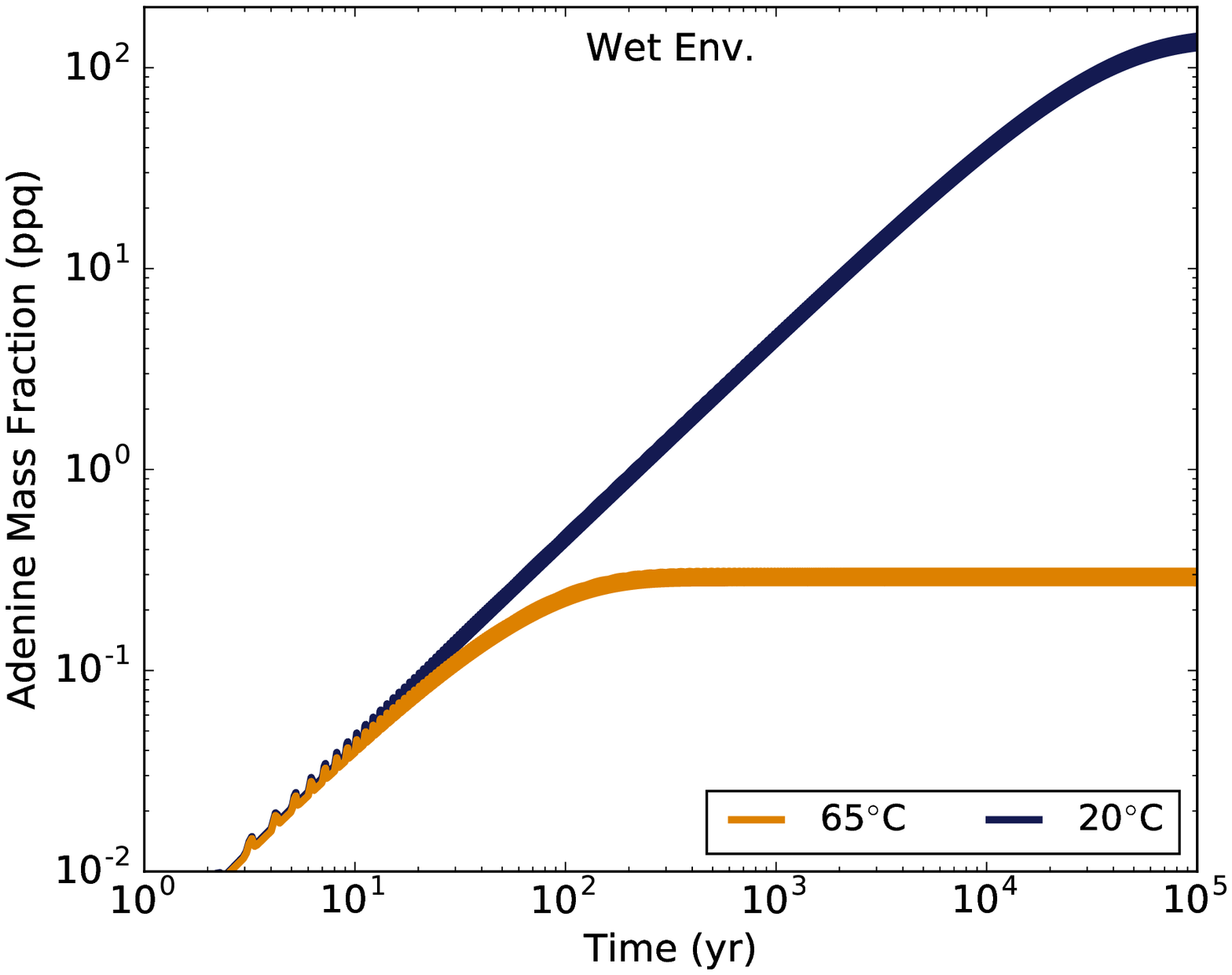}
\caption{{\bf The no seepage limit:} The accumulation of adenine from only carbonaceous IDP sources in cylindrical WLPs with radii and depths of 1 m. The two curves represent the adenine concentrations in a wet environment pond on a hot (65 $^{\circ}$C), and warm (20 $^{\circ}$C) early Earth (for details see Table~\ref{precipModels} in the main text). The thickness of the lines is due to the seasonal oscillations in adenine concentrations.}
\label{IDPAccumNoSeepage}
\end{figure}

In Figure~\ref{IDPAccumNoSeepage} we turn off seepage, e.g. resembling a scenario where a lipid biofilm has covered the WLP base, and explore the evolution of adenine concentrations from IDP sources. This model is displayed for a hot early Earth at 65 $^{\circ}$C and a warm early Earth at 20 $^{\circ}$C (See Table~\ref{precipModels} in main text for details). UV photodissociation is always the dominant nucleobase sink for the dry and intermediate environments, therefore for this model we only display the evolution of adenine in the wet environment, where hydrolysis takes over as the dominant nucleobase sink.

In the absence of seepage, adenine concentrations can build up in wet environment WLPs until the rate of incoming adenine from IDPs matches the decay rate due to hydrolysis. Hydrolysis rates are faster at hotter temperatures, therefore maximum adenine concentrations are higher and take longer to converge in the 20 $^{\circ}$C pond compared to the 65 $^{\circ}$C pond. However, these maximum adenine concentrations are 145 and 0.3 ppq, respectively, which are negligible in comparison to the ppb--ppm-level adenine concentrations reached in WLPs from carbonaceous meteorite sources.

\subsubsection*{Nucleobase Accumulation from Meteorites}

\begin{figure*}[htbp]
\centering
\includegraphics[width=17.8cm]{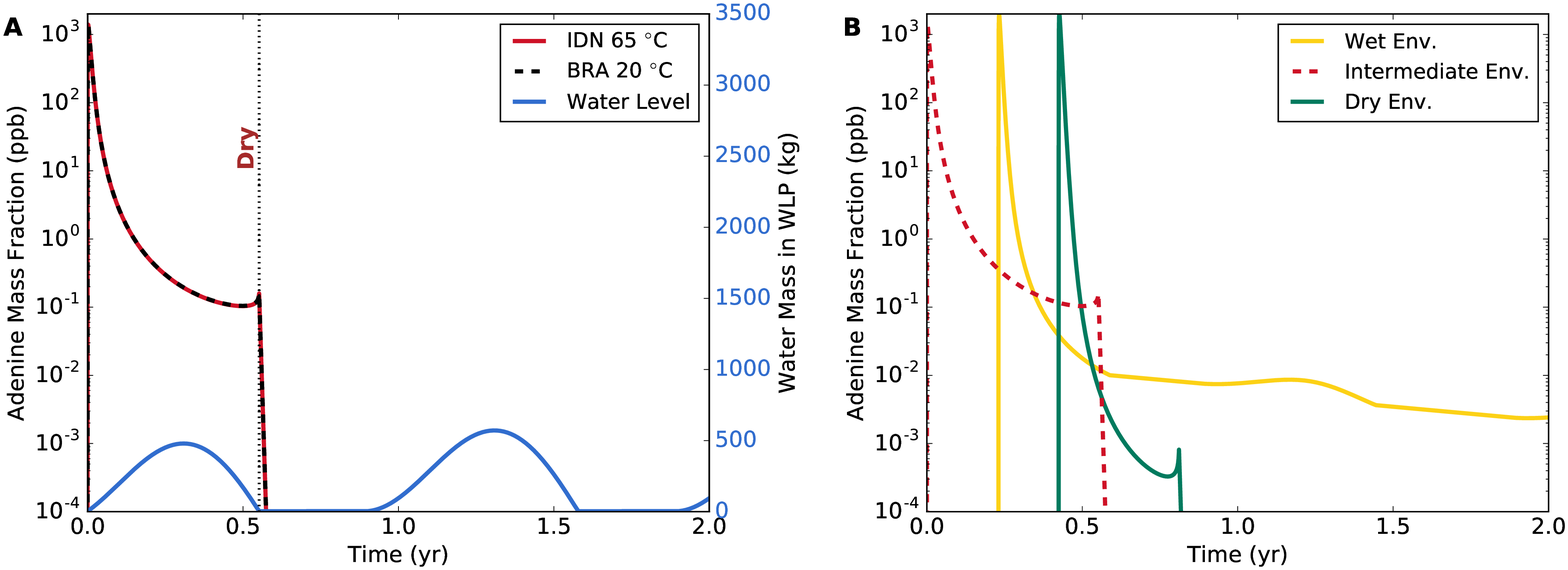}
\caption{The accumulation of adenine from 1 cm fragments of an initially 40 m-radius carbonaceous meteoroid in a cylindrical WLP with a radius and depth of 1 m. The degenerate WLP models used for these calculations correspond to a hot early Earth at 65 $^{\circ}$C and a warm early Earth at 20 $^{\circ}$C. {\bf (A)} The accumulation of adenine from carbonaceous meteorites in an intermediate environment (for details see Table~\ref{precipModels} in the main text). The wet-dry cycles of the pond are also shown to illustrate the effect of water level on adenine concentration. {\bf (B)} The three curves (dry, intermediate, and wet environments) differ by their precipitation rates, which are from a variety of locations on Earth today, and represent 2 classes of matching early Earth analogues: hot (Columbia, Indonesia, Cameroon), and warm (Thailand, Brazil, and Mexico) (for details see Table~\ref{precipModels} in the main text). The curves are obtained by numerically solving Equation~\ref{MetAccumEq}.}
\label{MetAccumAddWater}
\end{figure*}

In Figure~\ref{MetAccumAddWater} we explore the evolution of adenine concentration in WLPs with radii and depths of 1 m, from 1-cm fragments of an initially 40 m-radius carbonaceous meteoroid. The models correspond to degenerate environments on a hot (65 $^{\circ}$C) and warm (20 $^{\circ}$C) early Earth. The maximum adenine concentration in the intermediate environment is $\sim$1.4 ppm, and occurs 16 hours after the fragments deposit into the nearly empty pond. The dry and wet environments allow for a maximum adenine concentration of $\sim$2 ppm. Because adenine outflow from 1 cm-sized meteorites occurs in just 10 days, only adenine sinks exist from 10 days onwards. As the ponds wet, the adenine concentrations decrease exponentially, and as the ponds dry again, the curves flatten out and ramp up slightly before the ponds completely dry up. When the ponds dry, UV radiation quickly wipes out the adenine at the base of the ponds. Since the wet environment doesn't have a dry phase, the adenine concentration slowly diminishes in this model due mainly to seepage.

The adenine mass fraction curves in Figure~\ref{MetAccumAddWater}, over a 2 year period, do not change as pond radii and depths increase equally. This is because, although the mass of water in a WLP increases for larger collecting areas, larger pond areas also collect more meteorite fragments, which counterbalances the water mass and keeps the nucleobase concentration the same.

\begin{figure}[ht!]
\centering
\includegraphics[width=\linewidth]{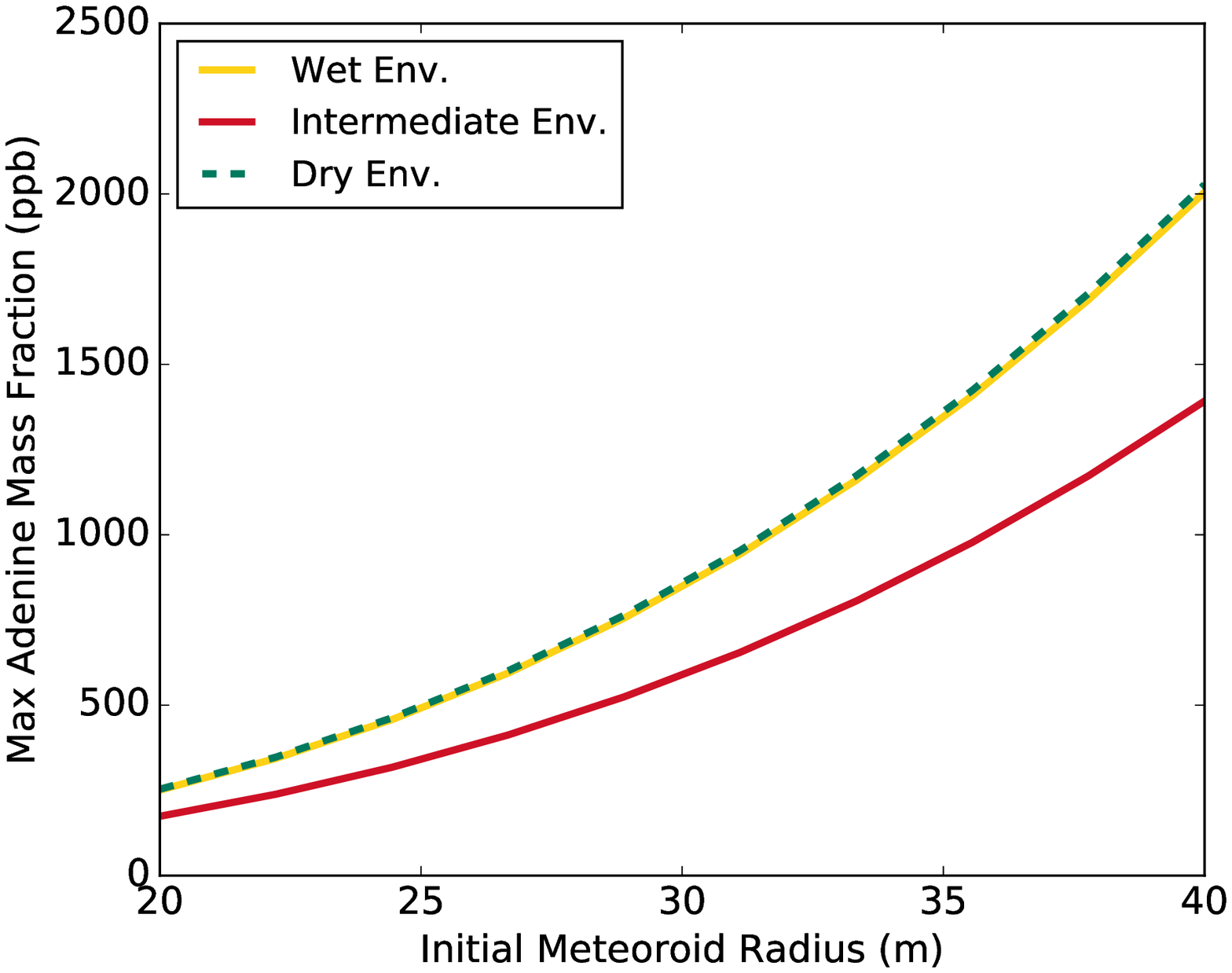}
\caption{The maximum concentration of adenine accumulated from 1 cm fragments of a carbonaceous meteoroid 20--40 m in radius in cylindrical WLPs with radii and depths of 1 m. The degenerate WLP models used for these calculations correspond to a hot early Earth at 65 $^{\circ}$C and a warm early Earth at 20 $^{\circ}$C. The three curves (dry, intermediate, and wet environments) differ by their precipitation rates, which are from a variety of locations on Earth today, and represent 2 classes of matching early Earth analogues: hot (Columbia, Indonesia, Cameroon), and warm (Thailand, Brazil, and Mexico) (for details see Table~\ref{precipModels} in the main text). The curves are obtained by numerically solving Equation~\ref{MetAccumEq}.}
\label{MetAccumByMetSize}
\end{figure}

In Figure~\ref{MetAccumByMetSize} we explore how initial meteoroid radius affects the maximum concentration of adenine accumulated (from its fragments) in WLPs with radii and depths of 1 m. The maximum adenine concentration only differs by at most a factor of 8 when varying the initial meteoroid radius from 20--40 m. This is because the nucleobase mass to enter a WLP scales with the meteoroid mass, i.e. $\propto$ r$^3$.

\begin{figure}[ht!]
\centering
\includegraphics[width=\linewidth]{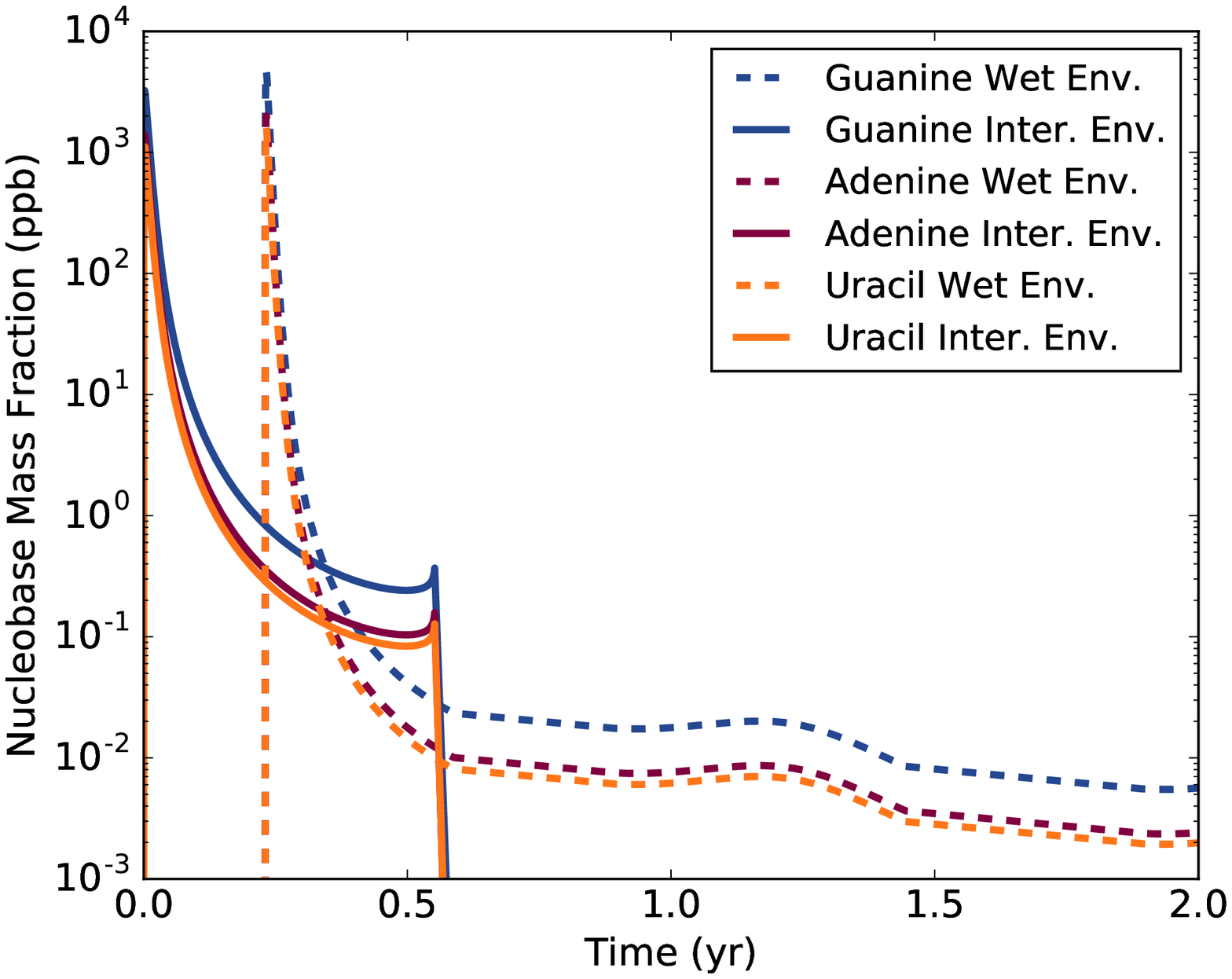}
\caption{The accumulation of guanine, adenine, and uracil from 1 cm fragments of an initially 40 m-radius carbonaceous meteoroid in cylindrical WLPs with radii and depths of 1 m. The degenerate WLP models used for these calculations correspond to a hot early Earth at 65 $^{\circ}$C and a warm early Earth at 20 $^{\circ}$C. The two curves for each nucleobase differ by their precipitation rates, which create intermediate (solid lines), and wet (dotted lines) environments, and are from a variety of locations on Earth today representing 2 classes of matching early Earth analogues: hot (Columbia, Indonesia), and warm (Thailand, Brazil) (for details see Table~\ref{precipModels} in the main text). The curves are obtained by numerically solving Equation~\ref{MetAccumEq}.}
\label{MetAccum65All}
\end{figure}

Finally, in Figure~\ref{MetAccum65All} we explore guanine and uracil accumulation in intermediate and wet environment WLPs with radii and depths of 1 m. These models correspond to a hot early Earth at 65 $^{\circ}$C and a warm early Earth at 20 $^{\circ}$C. The small differences between each nucleobase mass fraction over time is due to the different initial nucleobase abundances in the deposited meteorite fragments (see Table~\ref{AvgNucleoAbun}). Although each nucleobase has a different hydrolysis rate (see Equation~\ref{ade}), the decay of guanine, adenine, and uracil due to hydrolysis is negligible in $<$ 10 years for temperatures $<$ 65$^{\circ}$C.

\subsection*{Appendix A: Advection and Diffusion Model}\label{advecDiff}

Advection and diffusion are the two main considerations of solute transport in water. Because nucleobases will diffuse out of the pores of carbonaceous IDPs and meteorites at a different rate than they will mix homogeneously in the WLP, we separate our nucleobase transport model into two distinct parts. 
In part one, we model the outflow of nucleobases from carbonaceous IDPs and meteorites. In part two, we model the mixing of a local concentration of nucleobases into a WLP.

Both parts of our simulation can be modeled with the advection-diffusion equation below, with either one or both RHS terms ``turned on.''

\begin{equation}\label{fick}
\phi \frac{\partial C_i}{\partial t} = \nabla \cdot \left[ D_{eff} \nabla C_i \right] - \nabla \cdot  \left[u C_i \right],
\end{equation}
where $\phi$ is the porosity of the medium, $C_i$ is the mass concentration of the species, $D_{eff}$ is the effective diffusion coefficient (often in m$^2$s$^{-1}$), and $u$ is the convective fluid velocity \cite{2010E&PSL.296..235P}. 

For a 1D case, where the diffusion coefficient and fluid velocity are constant along the simulated path, the advection-diffusion equation can be written as, 

\begin{equation}\label{diffusioneqn}
\phi \frac{\partial C_i}{\partial t} = D_{eff}  \frac{\partial^2 C_i}{\partial r^2} - u \frac{\partial C_i}{\partial r}.
\end{equation}

For part one of our nucleobase transport model (the nucleobase outflow portion), we set $u$ = 0. This is because carbonaceous IDPs and meteoroid fragments are likely too small to attain noticeable interior pressure differences (thus the convective velocity within these bodies is probably negligible).

We do not consider hydrolysis in our nucleobase transport model, as we only intend on estimating nucleobase outflow and mixing timescales from these models (rather than the nucleobases remaining after these processes). Since nucleobase decay is uniform within the carbonaceous sources and WLPs, and is also very slow at WLP temperatures ($t_{1/2} \sim$ tens to hundreds of years \cite{Reference106}), it is not likely to affect the timescales of complete nucleobase outflow from the source or homogenization within the WLP. A non-zero amount of nucleobases will decompose during diffusion from the source, and during mixing within the WLP. However, this is considered in the final calculations of nucleobase accumulation within WLPs from each source (see Sections~\ref{IDPTheory} and \ref{MetTheory}).

The advection-diffusion equation also does not include adsorption or formation reactions. However, for the diffusion of {\it soluble} nucleobases from small porous environments which previously reached chemical equilibrium, the effects of these extra sources and sinks will probably be minimal. Also, to adjust the diffusion equation for a free-water medium, one simply needs to set $\phi$ = 1, and $D_{eff} = D_{fw}$.

The effective diffusion coefficient of a species is proportional to, but smaller than its free water diffusion coefficient. Many equations exist for modeling the effective diffusion coefficient in porous media \cite{Reference101,Reference102,Reference103,Reference104}. These equations depend on variables such as the porosity, tortuosity, and constrictivity of the medium, which represent the void space fraction, the curves in the pores, and the bottleneck effect, respectively. These equations are listed in Table~\ref{Deff}.

\begin{table*}[!ht]
\centering
\caption{Different equations for modeling the effective diffusion coefficient of species in porous media. Estimates of the effective diffusion coefficients of single nucleobases through the pores of carbonaceous IDPs and meteorites are also calculated for each model. The free water diffusion coefficient, $D_0$, represents the unobstructed diffusion of a species, the porosity factor, $\phi$, represents the void fraction of the medium, the constrictivity factor, $\delta$, represents the bottleneck effect due to small pore diameters, and the tortuosity factor, $\tau$, represents the restiction in diffusive flow due to curves in the pores. Estimates of these factors, and the empirical exponent, $m$, for carbonaceous IDPs and meteorites are, $D_0$ = 4$\times$10$^{-10}$m$^2$s$^{-1}$, $\phi$ = 0.25, $\delta$ = 1, $\tau$ = 1.45, and $m$ = 2 \cite{Reference105,2011M&PS...46.1842M,2013JGE....10e5003E,Reference104}. \label{Deff}}

\begin{tabular}{ccc}
\\
\multicolumn{1}{c}{$D_{eff}$} &
\multicolumn{1}{c}{Source} &
\multicolumn{1}{c}{Estimate for this work ($\times$10$^{-11}$ m$^2$s$^{-1}$)}\\ \hline \\[-2mm]
$\dfrac{\phi \delta}{\tau} D_{fw}$ & Saripalli $\emph{et al.}$ \cite{Reference101} & 6.90\\[4mm]
$\dfrac{\phi \delta}{\tau^2} D_{fw}$ & van Brakel and Heertjes \cite{Reference102} & 4.76\\[4mm]
$\dfrac{2 \phi}{3 - \phi} D_{fw}$ & Car{\'e} \cite{Reference103} & 7.27\\[4mm]
$\phi^m D_{fw}$ & Boving and Grathwohl \cite{Reference104} & 2.50\\
\hline
\multicolumn{3}{l}{\footnotesize $D_{fw}$ = free water diffusion coefficient}\\
\multicolumn{3}{l}{\footnotesize $\phi$ = porosity}\\
\multicolumn{3}{l}{\footnotesize $\delta$ = constrictivity}\\
\multicolumn{3}{l}{\footnotesize $\tau$ = tortuosity}\\
\multicolumn{3}{l}{\footnotesize $m$ = empirical exponent}
\end{tabular}
\end{table*}

Carbonaceous meteorites of type CM, CR, and CI have average porosities of 24.7$\%$, 9.5$\%$, and 35.0$\%$ \cite{2011M&PS...46.1842M}. Based on the relative fall frequency of these meteorites on Earth \cite{2002aste.book..653B}, the weighted average porosity of these meteorite types is $\sim$25$\%$. Chondritic IDPs have similar porosities to carbonaceous chondrites \cite{1997M&PS...32..509C}, therefore a 25$\%$ porosity may also well represent nucleobase-containing IDPs.

The constrictivity of a porous medium is only important when the size of the species is comparable to the diameter of the pores \cite{Reference104}. Therefore given that nucleobases are $<$ 1 nm in diameter and the bulk of pore diameters in, for example, the Acfer 094 carbonaceous chondrite, range from 20--200 nm \cite{2009E&PSL.287..559B}, we can neglect $\delta$ from the listed models.

Tortuosities of chondritic meteorites have an average value of 1.45 \cite{2013JGE....10e5003E}, and the empirical exponent $m$ for carbonaceous meteorites might be similar to that of nearshore sediments with a value of 2 \cite{Reference104}.

Finally, the free water diffusion coefficient of a single nucleobase has not been measured, however the free water diffusion coefficient of a single nucleotide is 400 $\mu$m$^2$s$^{-1}$ \cite{Reference105}. Since nucleotides are heavier than nucleobases by a ribose and phosphate molecule, they will likely diffuse slower than nucleobases. Therefore  400 $\mu$m$^2$s$^{-1}$ is a good estimate of the lower limit of the free water diffusion coefficient of a single nucleobase.

Using the above estimates, we calculate the effective diffusion coefficients for nucleobases in carbonaceous meteorites and IDPs using each of the four models and display them in their respective columns in Table~\ref{Deff}. The average value of the effective diffusion coefficient across all four models is 5.36 $\times$10$^{-11}$ m$^2$s$^{-1}$.

As previously stated, convective velocity within the pores of carbonaceous IDPs and meteorites is considered negligible. However, this is not the case within 1--10 m-radius WLPs. Due to the day-night cycles of the Earth, WLPs likely experienced a temperature gradient from the atmospherically exposed top of the pond to the constant geothermally heated base. Since convection is likely the dominant form of heat transport within hydrothermal ponds \cite{Reference107}, convection cells would have formed, with warm (higher pressure) parcels of water flowing upwards and recently cooled (lower pressure) parcels flowing downwards.

The convective fluid velocity can be estimated with the equation

\begin{equation}\label{convVel}
u = \sqrt{g \beta \Delta T L}
\tagaddtext{[m s$^{-1}$],}
\end{equation}
where $g$ is the gravitational acceleration experienced by the fluid (m s$^{-2}$), $\beta$ is the fluid's volumetric thermal expansion coefficient (K$^{-1}$), and $\Delta T$ is the temperature difference (K) over a scale length $L$ (m) \cite{Reference108}. The volumetric thermal expansion coefficients for water at 50, 65, and 80 $^{\circ}$C, are 4.7, 5.6, and 6.5 $\times$ 10$^{-4}$ K$^{-1}$ respectively \cite{Reference109}.

Small ponds and even lakes can experience a temperature difference of 1--5 $^{\circ}$C over the course of a day-night cycle \cite{Reference111}. However, convection begins cycling water well before temperature differences of this magnitude are reached. To estimate the lower bound of a constant temperature difference between the surface and the base of a WLP during a day-to-night period, we assume that each convection cycle cools a surface parcel of water by $\Delta T$. Then, given $\Delta T$, we match the corresponding cycle length and number of cycles with the approximate 12 hour period required for a 1 $^{\circ}$C change in pond temperature. The equation for this calculation is summarized below.

\begin{equation}\label{estimate1}
\Delta T = \left( \frac{2 L T_c}{P_{dn} \sqrt{g \beta L}} \right)^{\frac{2}{3}}
\tagaddtext{[K],}
\end{equation}
where $T_c$ is the change in pond temperature over a day-to-night period (K) and $P_{dn}$ is the duration of that period (s).

Using Equation~\ref{estimate1}, the minimum constant temperature difference between the surface and the base of a cylindrical WLP with a radius and depth of 1 m, at 65 $^{\circ}$C is $\sim$0.007 K. For a WLP with a radius and depth of 10 m, this minimum constant temperature difference increases to $\sim$0.016 K. However, given that smaller ponds experience greater temperature changes than larger ponds due to faster heat transfer, a $\Delta T$ of $\sim$0.01 K may be a reasonable lower-bound estimate for all WLPs in the 1--10 m-radius/depth range.

Given a constant temperature difference of 0.01 K between the base and surface of 1 m-, 5 m-, and 10 m-deep WLPs at around 65 $^{\circ}$C, the convective flow velocities would be approximately 0.7, 1.7, and 2.4 cms$^{-1}$, respectively.

Due to the 1D nature of our simulations, we are assuming a radially symmetric outflow of nucleobases from spherical carbonaceous IDPs and meteorites. We also assume that local concentrations of nucleobases recently flowed out of these sources will remain within a single convection cell. Lastly, we assume that the nucleobase homogenization timescale within a 1D convection cell of a WLP is mostly representative of the nucleobase homogenization timescale within the entire WLP. Though the 1D handling of this part of our model is a simplification of advection and diffusion within WLPs, since we are only attempting to estimate nucleobase homogenization timescales to within a few factors, a 1D model is probably sufficient. 

For both model parts we use a backward time, centered space (BTCS) finite difference method for the diffusion term in the advection-diffusion equation. For part two of the model, we use the upwind method to approximate the additional advection term. The BTCS method was selected over the more accurate Crank-Nicolson (CN) method based on the former's stability for sharply edged initial conditions and convergence for increasing levels of refinement. However, differences in diffusion timescales are found to be within rounding error upon comparison of these methods for a 100 $\mu$m carbonaceous IDP. The upwind method was selected over higher-order advection approximation methods (e.g. Beam-Warming, Lax-Wendroff) due to its lack of spurious oscillations, lowest error accumulation in mass conservation tests and convergence for increasing levels of refinement. These two models are summarized in Table~\ref{Models} below. 

\begin{table*}[!ht]
\centering
\caption{Summary of parts one and two of our 1D nucleobase transport model. Part one is a model of nucleobase outflow from carbonaceous IDPs and meteorites while they lay at the base of a WLP. Part two is the mixing (i.e. homogenization) of a local concentration of nucleobases throughout the WLP.\label{Models}}
\begin{tabular}{ccccccc}
\\
\multicolumn{1}{c}{Part} &
\multicolumn{1}{c}{Model Description} &
\multicolumn{1}{c}{Num. Method} &
\multicolumn{1}{c}{Boundaries} &
\multicolumn{1}{c}{Initial condition} &
\multicolumn{1}{c}{$\phi$} &
\multicolumn{1}{c}{$D_{eff}$ (m$^2$s$^{-1}$)}\\ \hline \\[-2mm]
1 &\begin{tabular}{@{}c@{}}Nucleobase outflow \\ from carbonaceous \\ IDPs and meteorites\end{tabular} & BTCS & \begin{tabular}{@{}c@{}}Neumann \\ and open\end{tabular} & See Figure~\ref{Init}a & 0.25 & 5.36 $\times$10$^{-11}$\\[6mm]
2 & \begin{tabular}{@{}c@{}}Nucleobase mixing \\ in WLPs\end{tabular} & \begin{tabular}{@{}c@{}}BTCS and\\ upwind\end{tabular} & Cyclic & See Figure~\ref{Init}b & 1 & 4.0 $\times$10$^{-10}$\\
\hline
\end{tabular}
\end{table*}

For part one of our nucleobase transport model, the simulation frame starts at the center of the IDP or meteorite, and ends at the rock-pond interface. The left ($r$ = 0) boundary is Neumann (i.e. $\dfrac{\partial C_i}{\partial t}$ = 0 across the boundary), simulating zero inflow, and the right boundary is open (i.e. $\dfrac{\partial C_i}{\partial t}$ before the boundary equals $\dfrac{\partial C_i}{\partial t}$ after the boundary), simulating outflow into the WLP. The nucleobase content of the modeled sources are initially homogeneous, but drop sharply to zero at the open boundary (i.e. the initial condition represents an exponentially smoothed step function). This is made to represent the searing of the outermost layer of carbonaceous IDPs and meteorites from atmospheric entry heating.

For part two, the simulation frame is an eccentric 1D convection cell, which loops between the bottom and the top of the WLP (length = 2$r_p$). The convection cell is modeled with cyclic boundaries, i.e., continuous nucleobase flow, and no nucleobases exit the convection cell. The initial concentration of nucleobases is a sharp Gaussian beginning at the base of the pond. The width of the spike is $\sim$1$\%$ of the WLP's radius.

\begin{figure*}[hbtp]
\centering
\includegraphics[width=17.8cm]{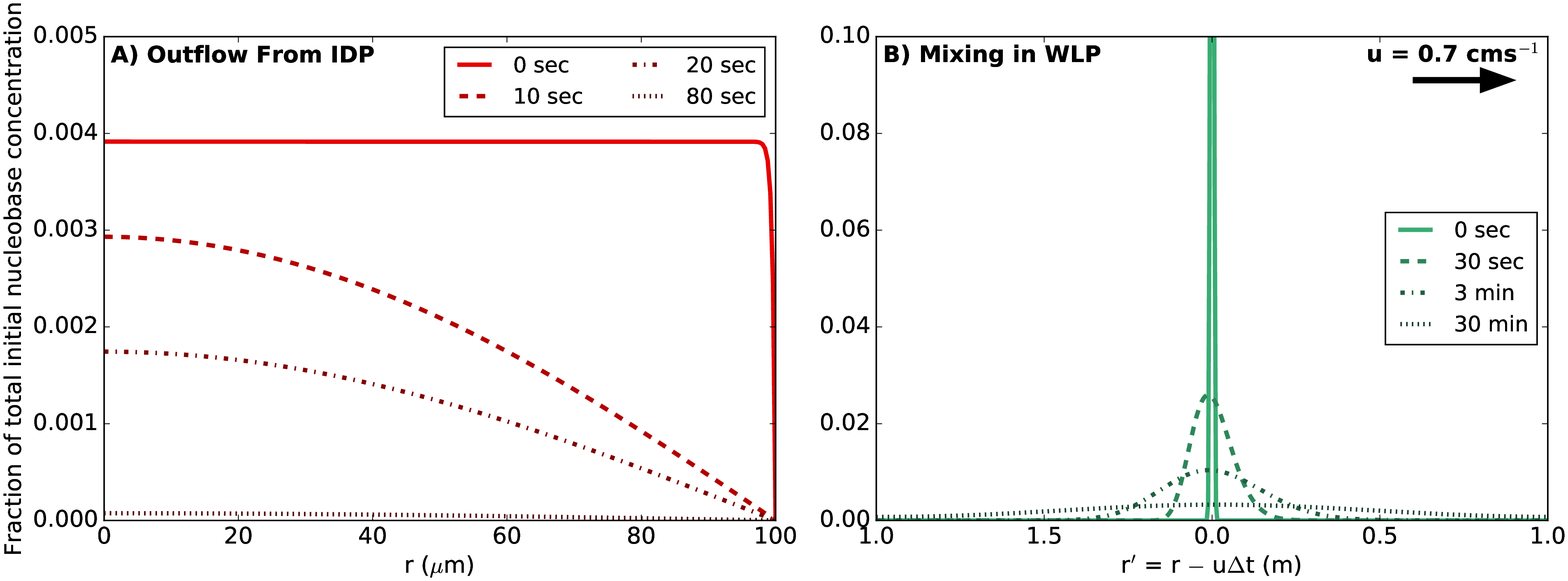}
\caption{Example simulations, including the initial conditions, for our two part nucleobase transport model. Each line represents a different snapshot in time. {\bf (A)} Initially homogeneous nucleobase diffusion from a 100 $\mu$m-radius carbonaceous IDP. {\bf (B)} Initially locally concentrated nucleobase mixing in a convection cell within a 1 m-deep WLP. The convection cell is a $L$ = 2$r_p$ eccentric loop flowing between the bottom to the top of the WLP. This loop is sliced at $r'$ = 1 m in the convection cell's moving frame and unraveled for display in the 1D plot in B.}
\label{Init}
\end{figure*}

Example simulations, including initial conditions, for parts one and two of our nucleobase transport model are plotted in Figure~\ref{Init}. Because the time it takes for complete nucleobase outflow from a source, or complete nucleobase homogenization in a WLP, is independent of initial nucleobase abundance, the solution to the diffusion equation at each grid point is displayed as a fraction of the total initial nucleobase concentration. For ease of viewing, part two of the model is displayed in the convection cell's moving frame, with coordinate $r'$ = $r$ - $u \Delta t$. 

\end{document}